\documentclass{aa}  

\usepackage{graphicx}
\usepackage{float}
\usepackage{txfonts}
\usepackage{multirow}
\usepackage[utf8]{inputenc}
\usepackage{natbib}
\usepackage{booktabs}
\usepackage{longtable}
\usepackage[flushleft]{threeparttable}
\usepackage{tabularx}
\usepackage{makecell}
\usepackage[fleqn]{amsmath}
\usepackage{hyperref}
\usepackage{ulem}
\usepackage{wrapfig} % Include the wrapfig package
\usepackage{upgreek} 
\hypersetup{
  colorlinks   = true,    % Colours links instead of ugly boxes
  urlcolor     = blue,    % Colour for external hyperlinks
  linkcolor    = blue,    % Colour of internal links
  citecolor    = blue      % Colour of citations
}
\usepackage[dvipsnames]{xcolor}
\bibpunct{(}{)}{;}{a}{}{,} % to follow the A&A style
\newcommand{\psr}{NGC~1851A}
\newcommand{\msun}{\rm M_{\odot}}
\newcommand{\stig}{\varsigma}
\newcommand{\MC}[1]{\textcolor{magenta}}
\newcommand{\rfifind}{\texttt{rfifind}}
\newcommand{\prepfold}{\texttt{prepfold}}
\newcommand{\presto}{\texttt{PRESTO}}
\newcommand{\psrchive}{\texttt{PSRCHIVE}}
\newcommand{\psradd}{\texttt{psradd}}
\newcommand{\tempo}{\texttt{TEMPO}}

\newcommand{\gptool}{\texttt{gptool}}

\newcommand{\simi}{$\sim$}

\newcommand{\us}{$\mu$s}

\newcommand{\mura}{$\mu_{\alpha}$}
\newcommand{\mudec}{$\mu_{\delta}$}
\newcommand{\masy}{mas\,yr$^{-1}$}
\newcommand{\Mp}{M_{\rm p}}
\newcommand{\Mc}{M_{\rm c}}
\newcommand{\bMp}{\bar{M}_{\rm p}}
\newcommand{\bMc}{\bar{M}_{\rm c}}
\newcommand{\mtot}{M_{\rm tot}}
\newcommand{\bMtot}{\bar{M}_{\rm tot}}

\defcitealias{Ridolfi_19}{R19}
\defcitealias{Joshi_Rasio_1997}{JR97}

\begin{document} 

\title{NGC~1851A: Revealing an ongoing three-body encounter in a dense globular cluster}

           \author{A.~Dutta \inst{\ref{mpifr}}\thanks{E-mail: adutta@mpifr-bonn.mpg.de}
           \and P.~C.~C. Freire \inst{\ref{mpifr}}
           \and T.~Gautam \inst{\ref{mpifr}, \ref{nrao}}
           \and N.~Wex \inst{\ref{mpifr}}
           \and A.~Ridolfi \inst{\ref{mpifr}, \ref{inaf}}
           \and D.~J.~Champion \inst{\ref{mpifr}}
           \and V.~Venkatraman~Krishnan \inst{\ref{mpifr}}
           \and C.~-H.~Rosie~Chen \inst{\ref{mpifr}}
           \and M.~Cadelano \inst{\ref{bologna1}, \ref{bologna2}}
           \and M.~Kramer \inst{\ref{mpifr}}
           \and F.~Abbate \inst{\ref{mpifr}, \ref{inaf}}
           \and M.~Bailes \inst{\ref{csiro}, \ref{ozgrav}}
           \and V.~Balakrishnan \inst{\ref{mpifr}}
           \and A.~Corongiu \inst{\ref{inaf}}
           \and Y.~Gupta \inst{\ref{ncra}}
           \and P.V.~Padmanabh \inst{\ref{aei},\ref{leibniz}}
           \and A.~Possenti \inst{\ref{inaf}}
           \and S.M.~Ransom, \inst{\ref{nrao}}
           \and L.~Zhang \inst{\ref{csiro}, \ref{naoc}} 
           % \and others \inst{\ref{inst2}}
          }

   \institute{Max-Planck-Institut f{\"u}r Radioastronomie, Auf dem H{\"u}gel 69, D-53121 Bonn, Germany\label{mpifr}
   \and
   National Radio Astronomy Observatory, 520 Edgemont Rd., Charlottesville, VA, 22903, USA\label{nrao}
   \and 
   INAF -- Osservatorio Astronomico di Cagliari, Via della Scienza 5, I-09047 Selargius (CA), Italy\label{inaf}
   \and
   Dipartimento di Fisica e Astronomia “Augusto Righi,” Universit\`a degli Studi di Bologna, 40129 Bologna, Italy.\label{bologna1}
   \and
   Osservatorio di Astrofisica e Scienze dello Spazio di Bologna, Istituto Nazionale di Astrofisica, I-40129 Bologna, Italy.\label{bologna2}
   \and 
   Centre for Astrophysics and Supercomputing, Swinburne University of Technology, P.O. Box 218, Hawthorn, VIC 3122, Australia\label{csiro} 
   \and
   Australian Research Council Centre of Excellence for Gravitational Wave Discovery (OzGrav)\label{ozgrav}
   \and 
   National Centre for Radio Astrophysics, Tata Institute of Fundamental Research, Pune 411007, Maharashtra, India\label{ncra}
   \and
   Max Planck Institute for Gravitational Physics (Albert Einstein Institute), D-30167 Hannover, Germany\label{aei}
   \and
   Leibniz Universit{\"a}t Hannover, D-30167 Hannover, Germany\label{leibniz}
   \and
   National Astronomical Observatories, Chinese Academy of Sciences, A20 Datun Road, Chaoyang District, Beijing 100101, People's Republic of China\label{naoc}
   }

   \date{Received --; accepted --}

\abstract{PSR J0514$-$4002A is a binary millisecond pulsar located in the globular cluster NGC~1851. The pulsar has a spin period of 4.99 ms, an orbital period of 18.8 days, and is in a very eccentric ($e = 0.89$) orbit around a massive companion. In this work, we present the updated timing analysis of this system, obtained with an additional 1 yr of monthly observations using the Giant Metrewave Radio Telescope and 2.5 yrs of observations using the MeerKAT telescope. Combined with the earlier data, this has allowed for the precise measurement of the proper motion of the system ($\mu_\alpha$ = 2.61 (13) mas $\mathrm{yr}^{-1}$ and $\mu_\delta$ = $-$0.90 (11) mas $\mathrm{yr}^{-1}$). This implies that the transverse velocity relative to the cluster is $30\,\pm\,7\,\mathrm{km}\,\mathrm{s}^{-1}$, which is smaller than the cluster's escape velocity, and thus consistent with the pulsar's association to NGC~1851. In addition to the spin frequency and its derivative, we have also confirmed the large second spin frequency derivative and large associated jerk (which has increased the spin frequency derivative by a factor of 27 since the mid-2000s). A measurement of the third spin frequency derivative for the pulsar showed that the strength of this jerk has increased by $\sim 65\%$ in the same time period. We have analysed the detailed implications of these measurements. First, we point out that to get a consistent picture of the orbital evolution, we must take the effect of the changing acceleration into account: this allows for much improved estimates of the orbital period derivative and solves one of the puzzles raised by previous timing. Second, we find that the large and fast-increasing jerk implies the presence of a third body in the vicinity of the pulsar. Based on our measured parameters, we constrain the mass, distance and orbital parameters for this third body. No counterpart is detectable within distance limit from \psr\, in the existing HST images. In any such configuration, the tidal contributions induced by the third body to the post-Keplerian parameters are relatively small, and the precise measurement of these parameters allowed us to obtain precise measurements of the total and component masses for the system: $\mtot = 2.4734(3) \, \msun$, $\Mp = 1.39(3)\,\msun$, $\Mc = 1.08(3)\,\msun$. This also indicates that the companion to the pulsar is a massive white dwarf and resolves the earlier ambiguity regarding its nature. Further observations will allow for the precise measurement of other higher frequency derivatives, allowing for the determination of the nature of the third body, and reveal whether it is gravitationally bound to the inner binary system. 
}


\maketitle

%%%%%%%%%%%%%%%%%%%%%%%%%%%%%%%%%%%%%%%%%%%%%%%%%%%%%%%%%%%%%%%%%%%%%%%%%%%%%%%%
   
\section{Introduction}

Globular clusters (GCs) are old, gravitationally-bound stellar systems that are roughly spherical in shape and contain some of the oldest stars in our galaxy. The large density of stars in their cores ($\sim$$10^{3}$ to $10^{6}\,\mathrm{pc}^{-3}$, \citealt{Baumgardt_18}) result in a high probability of close stellar encounters. Some of these encounters results in exchange interactions, where a main sequence (MS) star might exchange a MS companion with an old neutron star in the cluster. As the MS star evolves, it starts transferring matter to the NS, resulting in the formation of a low-mass X-ray binary (LMXB). For this reason, GCs are know to have $\sim 10^3$ times more X-ray binaries per unit of stellar mass than the Galactic disk \citep{Clark_1975}. This abundance is a clear sign that these X-ray binaries form dynamically.

In an LMXB, the NS is ‘recycled’ and ‘spun-up’ to high spin frequencies by accretion of matter from the evolving companion, leading to the formation of a millisecond pulsar (MSP) - a class of fully recycled pulsars with spin periods of a few milliseconds and very small spin period derivatives \citep{Tauris_Heuvel_2023}. In this process the orbit is circularised through tidal interactions. As the companion evolves to a white dwarf (WD) and the NS to an MSP, the low eccentricity of the system is normally retained, as observed in the vast majority of MSPs in the Galactic disk.

However, there are exceptions to this rule. In multiple star systems things can go very differently: One exceptional system observed in the Galactic disk is PSR~J1903+0327: this MSP ($P = 2.15$ ms) has a large orbital eccentricity ($e = 0.43$) and a massive companion \citep{Champion_2008}, which was later found to be a $1.03\,\msun$ main sequence star \citep{Freire_2011}. Detailed observational \citep{Freire_2011} and theoretical \citep{PZ_2011,Pijloo_2012} studies of this system indicate that it started its life as a triple system, which later became unstable, leading to the ejection of the donor to the pulsar.

Globular clusters are also multiple star systems. There are 344 pulsars currently known in 45 GCs\footnote{\href{https://www3.mpifr-bonn.mpg.de/staff/pfreire/GCpsr.html}{https://www3.mpifr-bonn.mpg.de/staff/pfreire/GCpsr.html}}; its MSP population represents about 40\% of the total known MSP population. Although many of these MSPs have circular orbits and WD companions, some GCs have a majority of isolated pulsars and slow pulsars with higher magnetic fields (e.g. Terzan 1, \citealt{Singleton_2024}, NGC~6517, \citealt{Yin_2024}, NGC~6522, \citealt{Abbate_2023}, NGC 6624, \citealt{Freire_2011,Abbate_2022}, NGC 6752, \citealt{Corongiu_2024}, M15, \citealt{Wu2024,Zhou2024}); additionally, some very dense GCs have MSPs with massive companions in eccentric orbits (e.g. Terzan 5, \citealt{padmanabh_2024}, NGC~6544, \citealt{Lynch_2012}, NGC~6624, \citealt{Ridolfi_2021}, NGC~6652, \citealt{DeCesar_2015} and M30, \citealt{Balakrishnan_2023}).

That these unusual pulsars are almost exclusively found in the denser clusters was explained by \cite{Verbunt+Freire} using the concept of the encounter rate per binary ($\gamma_\mathrm{i}$). A GC can have a large total stellar encounter rate and form many LMXBs and binary MSPs, but if the number of encounters {\it per star} over the age of the GC is small, then a LMXB will likely evolve undisturbed into a ``normal'' circular MSP - low-mass WD system, as generally observed in the Galactic disk and in the lower-density GCs. However, if the stellar density is very high, then the probability of subsequent encounters {\it for any particular system} increases. Only then are we likely to see a recycled pulsar - the end product of a LMXB formed in an exchange interaction - go into subsequent (``secondary'') exchange interactions. These can form exotic systems like the aforementioned MSPs with eccentric orbits and massive companions.

The GC NGC~1851 is located in the southern constellation of Columba at a distance of 11.66$\pm$0.25 kpc \citep{Libralato_2022}. It has a very high central density of about $3 \times 10^6 \rm \, M_{\odot} \, pc^{-3}$ \citep{Baumgardt_18} and has a moderately large encounter rate per binary ($\gamma_\mathrm{i} = 12.4\,\gamma_\mathrm{M4}$, \citealt{Verbunt+Freire}). The effect on the pulsar population is clear: of the 14 known pulsars, eight are in binaries \citep{Ridolfi_2022}; and of these, three have eccentric orbits and unusually massive companions \citep{Ridolfi_19,Barr_2024}, the characteristics of secondary exchange encounter products. These systems are especially interesting targets for further analysis.

One of these eccentric binaries, PSR~J0514$-$4002A, was found in 2003 using the Giant Metrewave Radio Telescope (GMRT). It was the first pulsar discovered in NGC~1851 \citep{Freire_2004}; for this reason it is also known (and will be designated in this work) as \psr. The timing solution was obtained from data taken with the Green Bank Telescope (GBT) \citep{Freire_2007} and later extended with data from the GMRT \citep{Ridolfi_19}. This binary pulsar is unlike anything found in the Galactic disk: its fast spin ($P \sim 5\,$ms), high orbital eccentricity ($e \sim 0.89 $) and large  ($\sim 1 \, \rm M_{\odot}$) companion mass made it the second clear case of a secondary exchange encounter product (the first case was PSR~B2127+11C in the globular cluster M15, see \citealt{Prince_1991}). 

In this work, we aim to address the unresolved questions about \psr\, raised by \citealt{Ridolfi_19} (hereafter \citetalias{Ridolfi_19}). First, their estimate of the proper motion suggested that the pulsar was not bound to the cluster. Secondly, the  observed large positive change in the orbital period of the system could not be explained by the acceleration caused by the gravitational field of the GC, as this would also cause a large positive spin period derivative, which is not observed. Third, their mass measurements left open the question about the nature of the companion. Finally, and most importantly, they measured a large second spin frequency derivative, which is indicative of a very large change of acceleration of the system (a ``jerk''), but the implications of this were not discussed in detail.

These puzzles motivated an extension of the timing dataset with sensitive observations from the MeerKAT radio telescope over a long time period, with a few dedicated observations of the system around the periastron and superior conjunction. This extended timing analysis allows for a robust measurement of the proper motion, the orbital period derivative, improvements of the constraints on the other relativistic parameters of the system, which in turn allow more precise measurements of the component masses and a better measurement of the jerk and its change with time. 

In this paper, we present the up-to-date timing solution of \psr, obtained with observations made using the upgraded GMRT (henceforth uGMRT) and MeerKAT, in addition to the earlier GBT and GMRT dataset from \citetalias{Ridolfi_19}, over a total time span of \simi 18 years. The observations, the subsequent data reduction, and how the resulting pulse times of arrival (ToAs) were analysed are discussed in Section~\ref{sec:observations}. The system parameters and other derived results benefit largely from the much longer timing baseline and the inclusion of new datasets with good timing precision; these are discussed in Section~\ref{sec:results}. Apart from the timing parameters of \psr, this section addresses the unresolved issues from \citetalias{Ridolfi_19}. Next comes the analysis of the large observed jerk, which we argue is caused by the presence of a third object near the \psr\, system. In Section~\ref{sec:period derivatives}, we discuss the implications of this jerk for understanding the variation of the orbital period and the spin-down of the pulsar. In Section~\ref{sec:external influence}, we explore the characteristics of this third object. In Section~\ref{sec:chi2map}, we update the mass measurements for the \psr\, system. In Section~\ref{sec:conclusions} we briefly summarise the results and arrive at a few general conclusions about the nature of the system.

%--------------------------------------------------------------------
\section{Observations and data analysis}
\label{sec:observations}

The new data analysed in this work include the observations taken with the uGMRT between 2018 and 2020 and the ones that were later carried out using MeerKAT, spanning over a total time of more than 4 years. A full description of the dataset, including a breakdown of the individual datasets is given in Table \ref{table:Observations}.

%Observations with GMRT
\subsection{GMRT Observations}
The observing campaigns for \psr\ with the GMRT have been running since 2017 April (MJD 57872), during which the data were taken in the Phased Array (PA) mode with a sampling time of 81.92 \us. Since September 2017, besides the PA mode, all observations are also taken simultaneously using the coherent de dispersion (CDP) mode of the GMRT Wideband Backend \citep{De_2016, Reddy_2017}. For all observations made in the CDP mode, the data were coherently de dispersed at the dispersion measure (DM) of \psr\ of $\sim$$52.14$ pc cm$^{-3}$; for a detailed description of these data, see \citetalias{Ridolfi_19}. This work looks into all CDP data taken for \psr\ between 2019 April (MJD 58601) to 2020 March (MJD 58928) with the 250-500 MHz Band-3 receiver of the uGMRT, a substantially longer dataset than had been analysed by \citetalias{Ridolfi_19} (Table~\ref{table:timing_params}).

%Observations with MeerKAT
\subsection{MeerKAT Observations}
NGC 1851 has been observed by MeerKAT, as part of both the MeerTime \citep{Bailes_2020} and TRAPUM \citep{Stappers_Kramer_2016} large science programs. Most of MeerKAT data we use in this paper were taken as part of the MeerTime GC pulsar timing programme. We have observed the pulsar with the MeerKAT on 27 different epochs, from January 2021 to April 2023 (Table~\ref{table:Observations}), and the data were recorded in coherently dedispersed, full-Stokes mode using the PTUSE backend \citep{Bailes_2020}. The first seven timing observations with MeerTime, which targeted precisely \psr, were taken in parallel to the observations performed for TRAPUM. This includes 3 observations done with the L-band receiver at a central frequency of 1284 MHz and a bandwidth of 856 MHz, and the rest using the UHF receivers at a central frequency of 816 MHz and a bandwidth of 544 MHz. These data were sampled every 7.53 $\upmu$s and possessed 4096 channels.

Further UHF observations, carried out during the MeerKAT GC survey, were recorded with a sampling time of 9.41~\us\ across 1024 channels, each with a bandwidth of $\sim$ 0.53 MHz. Later, the sampling time was increased to 15.06 \us\ to reduce the acquired data volume. The different channelisations of the UHF data resulted in a 0.2 MHz difference in the central frequencies. All observations taken with MeerKAT ranged between 2 hours to 4 hours in duration, amounting to a total of 64 hours, and the details for all observations are presented in Table~\ref{table:Observations}. A detailed description of the setups used for these observations can be found in \cite{Ridolfi_2022} and \cite{Barr_2024}.

% %Observations table
\renewcommand{\arraystretch}{1.5}
\begin{table*}[!h]
\setlength\tabcolsep{4.2pt}%
 \caption{Details of observations of \psr\ taken with the uGMRT and the MeerKAT. $t_{\rm samp}$: sampling time, $f_{\rm c}$: central frequency, $\Delta$f: observing bandwidth, $N_{\rm chan}$: number of frequency channels. We have used, in addition, the earlier
 GBT dataset taken by \cite{Freire_2007} and \cite{Ridolfi_19}.}
 \label{table:Observations}
 \centering
  \begin{tabularx}{\textwidth}{ccccccccccc}
  \hline
  Telescope & Mode/ & Receiver & Number of & Span & $t_{\rm samp}$ & $f_{\rm c}$ & $\Delta$f & $N_{\rm chan}$ & $\#$TOAs & EFAC\\
       & Backend & & observations & (MJD) & ($\mu$s) & (MHz) & (MHz) &  \\
 \hline
 \multirow{2}{*}{uGMRT} &CDP &  Band 3 & 12 & 58601$-$58867 & 10.24 & 399.80 & 200 & 512 & 203 & 1.76 \\
%  \\\cline{2-10} &&&
  & PA & Band 3 & 2 & 58652,58928 & 81.92 & 399.80 & 200 & 2048 & 13 & 1.74\\
  \hline 
  \multirow{2}{*}{MeerKAT} & {L-band} & {PTUSE} & 3 & 59229$-$59342 & 9.57 & 1284 & 856 & 4096 & 16 & 0.96\\
  & {UHF} & {PTUSE} & 24 & 59355$-$60047 & 7.53$^*$/9.41 & 816 & 544 & 4096$^*$/1024 & 825 & 1.22\\
    \hline
  \end{tabularx}
  \begin{tablenotes} 
    \item $^*$ piggybacked MeerTime data for TRAPUM observations that use the 4096 channelisation mode.
  \end{tablenotes}
\end{table*}

%--------------------------------------------------------------------
%--------------------------------------------------------------------
%--------------------------------------------------------------------

%GMRT
\subsection{Data analysis}

% \cM{GMRT I:}
For all GMRT observations taken in the PA and the CDP mode with 200 MHz bandwidth, offsets of 1.34217728 s and 2.01326592 s \footnote{Reddy et al, \href{http://www.gmrt.ncra.tifr.res.in/subsys/digital/DigitalBackend/target_files/GWB/ITR/GWB_Timestamp_ITR_Release_1.0.pdf}{NCRA Internal Technical Report}, April 2022} were included in the analysis for the respective set of TOAs (Appendix~\ref{sec:appendix_GMRT_clock}). Given the 4.99 ms period of this pulsar, these corrections accounted for hundreds of rotations, which would be missed if these offsets were not taken into account.

% \cM{GMRT II:gptool, rfifind, and folding:} 
The initial step of analysis for all search-mode observations taken with the uGMRT was done using \gptool\footnote{\url{https://github.com/chowdhuryaditya/gptool}}, a tool used for data reduction and bandpass normalisation of the beamformer data of the uGMRT \citep{Susobhanan_2021}. The ephemeris obtained from the analysis of \citetalias{Ridolfi_19} was used to fold the pulsar at its topocentric period for every observation, after an initial rfi removal, using the \prepfold\ and \rfifind\ routines of \presto\footnote{\url{https://github.com/scottransom/presto}} \citep{Presto} respectively. These folded archives were then converted into PSRFITS format, using the \texttt{psrconv} routine of \texttt{PSRCHIVE}, and the number of frequency channels in the resulting folded archives were downsampled to 256. For the two frequency bands, a high signal-to-noise ratio analytic profile (a ``template'') was created by adding up all the archives using the \psradd\ routine of \psrchive\footnote{\url{http://psrchive.sourceforge.net/}} and fitting two Gaussian components to it. 

% \cM{Analysis}: 
The pulse profiles for each of the detections of the pulsar were then cross-correlated with the corresponding templates, and TOAs were extracted using the \texttt{pat} routine of \psrchive. Depending on the scintillation and the brightness of the pulsar for individual observations, TOAs were made every \simi 10-60 minutes, and with 2 and 4 sub-bands for the GMRT and the MeerKAT datasets respectively. In addition to the offsets for the GMRT datasets described above, a time difference of 1.88 $\mu\mathrm{s}$ was also included for the TOAs created from the latter UHF archives to account for the 0.2 MHz difference in the central frequency\footnote{This was done using the TIME statement from TEMPO}. This corresponds to the difference in the sampling time between the two sets of the UHF observations. A careful phase alignment of the standard profiles for the CDP observations of the GMRT and the L-band and UHF observations of MeerKAT was done to a phase precision of $0.01$ for the removal of any arbitrary timing offset \citep{Guo_2022}. 
%--------------------------------------------------------------------

\subsection{Timing Procedure}
% \cM{Adding and re-doing analysis from \citetalias{Ridolfi_19}:}
In this work, the observations taken with MeerKAT and uGMRT were added to the earlier data used by \citetalias{Ridolfi_19}. For the CDP data from the GMRT discussed in \citetalias{Ridolfi_19}, all the observations were re-folded after performing the standard procedure of bitshift conversion (down-sampling the data from a 16-bit to a 8-bit format, \citealt{Gautam_2022}), bandpass normalisation and rfi mitigation, as is used for the rest of the uGMRT dataset as described above. This procedure has enabled the use of a single template for all of the GMRT CDP data and ensured a better timing precision. The final addition of the latest observations taken in April 2023 with MeerKAT has extended the timing baseline of the pulsar to almost two decades.

% \cM{Timing analysis procedure:}
We have used the \tempo\footnote{\url{http://tempo.sourceforge.net/}} timing package for the analysis of the TOAs and the Jet Propulsion Laboratory's DE 421 Solar System ephemeris \citep{Folkner_2009} to take into account the motion of the radio telescopes relative to the Solar System Barycentre (SSB). The parameters in the timing solution are specified in dynamical barycentric time (TDB). 

% \cM{DDGR and DDH models:} 
For the initial description of the orbit we used the DDH and the DDGR orbital models, which are derived from the ``DD'' orbital model \citep{DD_1986}. Like the DD model, the DDH model is a theory-independent model, where each relativistic effect is quantified  by a ``post-Keplerian'' parameter. The main difference relative to the DD model is that it uses the orthometric parametrisation of the Shapiro delay \citep{Freire_Wex}; this was chosen because the two post-Keplerian parameters that describe the Shapiro delay ($h_3$ and $\varsigma$) are much less correlated than the ``range'' and ``shape'' parameters that describe the Shapiro delay in the DD model (especially for orbital inclinations that are far from edge-on, as is the case for \psr).
These parameters can be related to the Keplerian parameters and the masses of the components of the system using a gravity theory. 
The DDGR model is a modification of the ``DD'' model that fits directly for the total mass and the companion mass of the system under the assumption that all relativistic effects are as predicted by general relativity (GR).

% \cM{Using DDH and DDGR models:} 
The significant change of the line-of-sight acceleration (the jerk) inferred from the spin period derivatives required the inclusion of higher order orbital frequency derivatives for the system (see further details in section ~\ref{sec:period derivatives}). In principle, this can be done easily using the BTX model \citep{Shaifullah_2016}, which is a re-implementation of the BT model \citep{BT_1976} that includes higher order derivatives for the orbital frequency. However, the BTX model is not suitable for eccentric systems like \psr\ since it estimates that the longitude of periastron increases linearly with time. The DD-based timing models, on the other hand, account correctly for the evolution of the longitude of periastron, which increases linearly with the true anomaly. This feature significantly improves the quality of the fit for \psr. However, as currently implemented in {\sc tempo}, these models are unable to account for higher orbital frequency derivatives. For this reason, we implemented modified versions of these models that can take into account the jerk of the system in a self-consistent way (see details in Appendix~\ref{sec:DD_modification}). All updated parameters and results reported in this work have been derived using these modified DD models.

%%%%%%%%%%%%%%%%%%%%%%%%%%%%%%%%%%%
%%%%%%%%%%%%%%%%%%%%%%%%%%%%%%%%%%%

\section{Results}
\label{sec:results}
%%%%%%%%%%%%%%%%%%%%%%%%%%%%
%%%%%%%%%%%%%%%%%%%%%%%%%%%%

\begin{figure*}[h!]
    \centering
    \includegraphics[width=\linewidth]{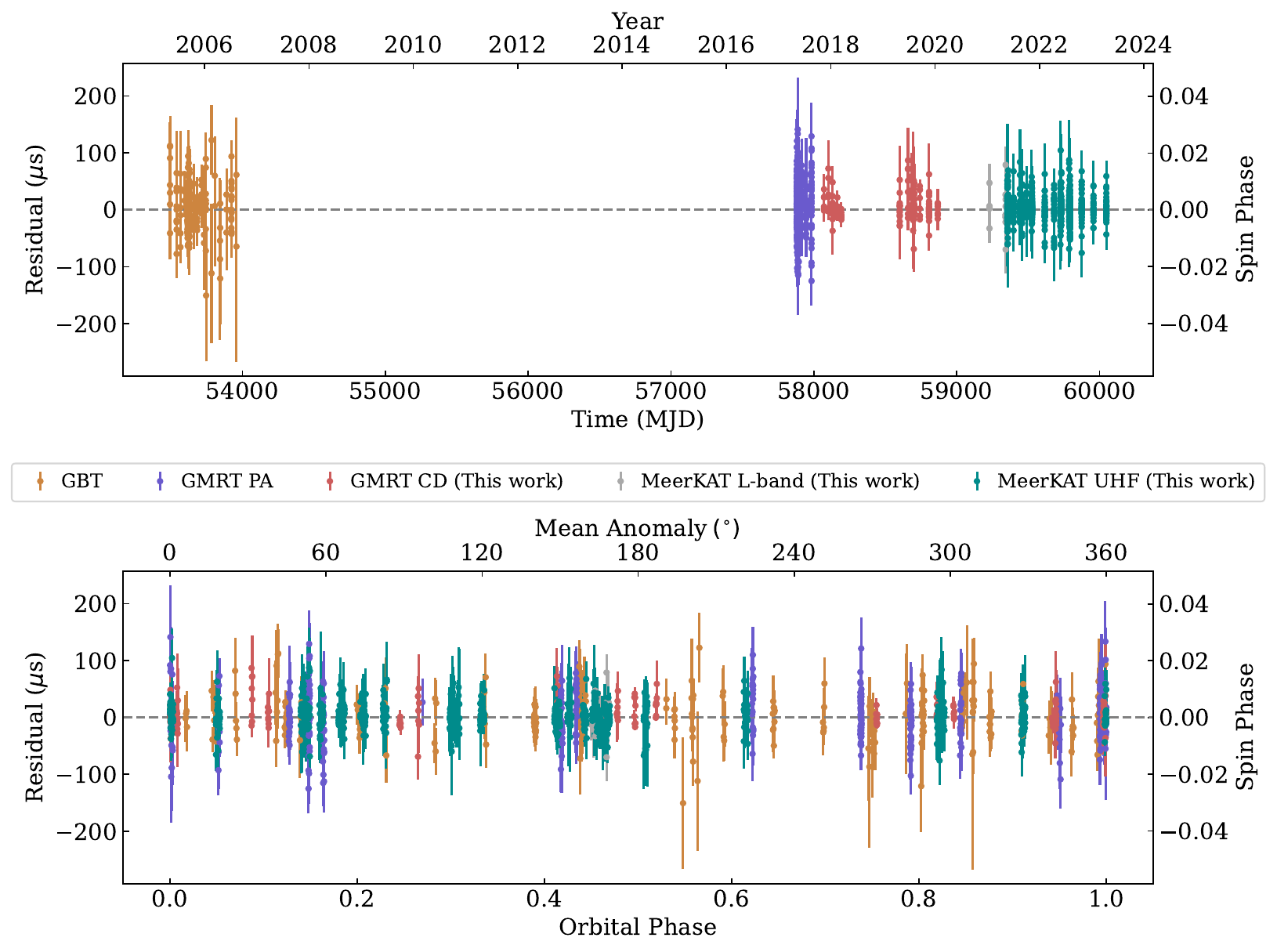}
    \caption{Top to bottom: Post-fit timing residuals for \psr\ obtained with the DDH timing model, as a function of (a) epoch and (b) orbital phase. The orbital phase 0 denotes periastron and the residual 1-$\sigma$ uncertainties are indicated by vertical error bars. The various colours denote the different telescopes and backends used for the timing. \textit{Orange} indicates the early GBT data, and \textit{purple} and \textit{red} the uGMRT data taken in the PA and CDP modes respectively. \textit{Grey} and \textit{green} represent the data taken with the L-band and the UHF receivers of the MeerKAT.}
    \label{1851A_Residuals}
\end{figure*}

The extended timing baseline of $\sim$18 years, inclusion of the 2-s offset in the GMRT dataset (see Appendix~\ref{sec:appendix_GMRT_clock}) and modified analysis procedure has been the key in obtaining precise measurements of a handful of astrometric and spin parameters for \psr. These values and their interpretations are discussed in Sect.~\ref{sec:results}. The updated timing parameters for \psr\ obtained after this analysis are presented in Table~\ref{table:timing_params}. All the parameters and their uncertainties are derived using the TEMPO timing software and the modified DDH model mentioned above. 
\\
The residuals between the predicted and measured times of arrival of the pulsar, as a function of time and orbital phase, are given in Fig.~\ref{1851A_Residuals}. There are no apparent systematic trends in the residuals, and the reduced $\chi^2 = 1.0044$ of the overall fit shows that the timing model provides a good estimation of the TOAs. This was obtained after individual adjustments for each of the GBT, GMRT, and MeerKAT datasets separately, and these adjustment factors (EFACs) are noted in Table~\ref{table:Observations}. 

\begin{table}[!h]
%\large
\caption{Timing solution for \psr, derived using {\sc tempo} with the modified DDH binary model. The values in the parenthesis indicate 1-$\sigma$ uncertainties on the last digit. 
% Derived parameters in cyan are derived using the DDGR model, updated PEPOCH.
}
\centering 
\begin{scriptsize}
\begin{tabular}{lc}
\hline
\hline
\multicolumn{2}{l}{Dataset and fixed quantities}\\
\hline
Terrestrial Time Standard   \dotfill &  TT(BIPM2019)  \\
Time Units \dotfill    & TDB  \\
Solar System Ephemeris  \dotfill &  DE421  \\
Span of timing data (MJD) \dotfill & 53493 -- 60047\\
Reference Epoch (MJD)\dotfill & 59728.338327\\
Number of TOAs \dotfill & 1791\\
Weighted rms residual (${\upmu} \rm s$) \dotfill & 12.54\\
$\chi^2$  \dotfill & 1772.70 \\
Reduced $\chi^2$ \dotfill & 1.0044 \\
Parallax (mas) \dotfill & 0.0857 \\
\hline
% \multicolumn{2}{l}{Spin and astrometric parameters}\\
\multicolumn{2}{l}{Timing parameters} \\
\hline
Right ascension, $\alpha$ (J2000)\dotfill & 05:14:06.697099(37) \\
Declination, $\delta$ (J2000)\dotfill & $-$40:02:48.90709(23) \\
Proper motion in $\alpha$, $\mu_{\alpha}$ (\masy)\dotfill & $2.61(13)$  \\
Proper motion in $\delta$, $\mu_{\delta}$ (\masy)\dotfill  & $-$0.90(11) \\
Spin frequency, $\nu$ (Hz)\dotfill & $200.37770716645(1)$ \\
First spin frequency derivative, $\dot{f}$ ($10^{-16}$\,Hz\,s$^{-1}$)\dotfill & $-$9.451(8)  \\
Second spin frequency derivative, $\ddot{f}$ ($10^{-24}$\,Hz\,s$^{-2}$)\dotfill & $-$2.14(2)  \\
Third spin frequency derivative, $\dddot{f}$ ($10^{-33}$\,Hz\,s$^{-3}$)\dotfill &$-$1.57(8)  \\
Dispersion measure, DM (cm$^{-3}$\,pc) \dotfill  &  52.1346(2) \\
First Derivative of DM, DM1 ($10^{-3}\,\rm cm^{-3}\,pc\,yr^{-1}$)\dotfill  & $-$1.2(2) \\
Second Derivative of DM, DM2 ($10^{-5}\,\rm cm^{-3}\,pc\,yr^{-2}$)\dotfill  & 2.2(1) \\
Third Derivative of DM, DM3 ($10^{-5}\,\rm cm^{-3}\,pc\,yr^{-3}$)\dotfill  & 2.5(19) \\
% \hline
% \multicolumn{2}{l}{Orbital parameters}\\
% \hline
Orbital period, $P_\mathrm{b}$ (day)\dotfill & 18.785179196(2) \\
Orbital eccentricity, $e$ \dotfill & 0.8879765(3) \\
Projected semi-major axis of pulsar orbit, $x \equiv a_\mathrm{p}\sin i/c$ (s) \dotfill &  36.2909(1) \\
Epoch of periastron, $T_{0}$ (MJD) \dotfill & 59728.33832689(13) \\
Longitude of periastron, $\omega \, (\deg)$ \dotfill & 82.548(2) \\
Rate of advance of periastron, $\dot\omega \, (\mathrm{deg}\,\mathrm{yr}^{-1})$ \dotfill & 0.012961(1) \\
Einstein delay, $\gamma$ (s) \dotfill & 0.01925(48) \\
Observed orbital period derivative, $\dot{P}_{\mathrm{b}}$ (10$^{-12} \mathrm{s}\,\mathrm{s}^{-1})$ \dotfill & 2.2(7) \\
Orthometric amplitude of Shapiro delay, $h_{3} (\upmu\mathrm{s})$ \dotfill & 0.52(53) \\
\hline
\multicolumn{2}{l}{Derived parameters}\\
\hline
Offset from the centre of the cluster (arcsec) \dotfill & 1.58 \\
Total proper motion, $\mu_{\text{T}}$ (\masy) \dotfill  & 2.76(12) \\
Position angle of proper motion, J2000, $\Theta_{\mu}$ ($\deg$) \dotfill &  109(2) \\
Proper motion relative to the cluster (\masy) \dotfill  &  0.54(13)\\
Transverse velocity relative to the cluster (km s$^{-1}$) \dotfill  &  30(7) \\
Spin period, $P$ (ms) \dotfill & $4.9905751200622(4)$ \\
Spin period derivative, $\dot{P}_\mathrm{obs}$ ($10^{-20} \mathrm{s}\,\mathrm{s}^{-1}$) \dotfill &2.354(2)  \\
Second orbital frequency derivative, $\ddot{f}_\mathrm{b} (10^{-33}$ Hz s$^{-2}$) \dotfill & $-$6.58 \\
Third orbital frequency derivative, $\dddot{f}_\mathrm{b} (10^{-42}$ Hz s$^{-3}$) \dotfill & $-$4.84 \\
Intrinsic spin period derivative, $\dot{P}_\mathrm{int}$ ($10^{-20}\, \mathrm{s}\,\mathrm{s}^{-1}$) \dotfill &  1.632(2)\\
Characteristic age (Gyr) \dotfill  & $4.8$ \\
Surface magnetic field ($10^8$ G) \dotfill & 2.8\\
Mass function $(\msun)$ \dotfill & 0.145428(2) \\
Total system mass, $\mtot\ (\msun)$ \dotfill & 2.47346(28) \\
Orbital inclination, $i$ (deg) \dotfill &  63 (3)\\
Pulsar mass, $\Mp \ (\msun)$ \dotfill & 1.39(3) \\
Companion mass, $\Mc\ (\msun)$  \dotfill & 1.08(3) \\
GW orbital decay, $\dot{P}_{\mathrm{b,int}}$ (10$^{-12} \mathrm{s}\,\mathrm{s}^{-1})$ \dotfill & $-$0.153 \\
\hline
\label{table:timing_params}
\end{tabular}
\end{scriptsize}
\end{table}

%%%%%%%%%%%%%%%%%%%%%%%%%%%%
%%%%%%%%%%%%%%%%%%%%%%%%%%%%

%%%%%%%%%%%%%%%%%%%%%%%%%%%%%%%%%%%%%%%%%%%%%
\subsection{Astrometric parameters}
We find that the position of the pulsar at the reference epoch is $\alpha \, = \, 05^{\mathrm{h}}14^{\mathrm{m}}06\fs697099(37)$ and $\delta \, = \,-40^{\circ}02^{'}48\farcs90709(23)$, which is 1\farcs58 from the centre of the GC at $\alpha^{'}\, = \, 05^{\mathrm{h}}14^{\mathrm{m}}06\fs755$ and $\delta^{'} \, = \,-40^{\circ}02^{'}47\farcs47$ \citep{Miocchi2013}. The latter study also includes an estimate of the core radius, 5\farcs4, this means that the pulsar is well within the core of the cluster.

The quantity that has changed the most notably from the introduction of the additional $2$s offset in the GMRT data and the continued timing analysis, is the proper motion of the system. We now obtain \mura = 2.61 $\pm$ 0.13 \masy\ in right ascension and \mudec = $-$0.90 $\pm$ 0.11 \masy\ in declination. The total proper motion of \psr\ is 2.76\,(12) \masy\ and the corresponding position angle is $109\,(2)\,\deg$. Using the images obtained from the Hubble Space Telescope (HST) Legacy Survey of Galactic Globular Clusters and GAIA astrometric data, \cite{Libralato_2022} have provided an updated measurement of the proper motion of the host cluster in right ascension and declination as $\mu_{\alpha} = 2.128 \pm 0.031 \, \rm mas \, yr^{-1}$, $\mu_{\delta} = -0.646 \pm 0.032 \, \rm mas \, yr^{-1}$. This means that the motion of the pulsar relative to the cluster is $\Delta\mu_{\alpha} =  0.48 \pm  0.13\, \rm mas \, yr^{-1}$, $\Delta\mu_{\delta} =  -0.25 \pm 0.11\, \rm mas \, yr^{-1}$. The corresponding relative transverse velocity of \psr\, to the centre of the cluster, using the distance estimate of $d = 11.66 \pm 0.25 \, \mathrm{kpc}$ \citep{Libralato_2022}, is $v_\mathrm{T} = 30 \pm 7 \, \mathrm{km}\,\mathrm{s}^{-1}$. While the three-dimensional velocity for any binary pulsar in a GC is not known, the transverse velocity of \psr\ is smaller than the cluster's central escape velocity ($v_\mathrm{esc}$) of $42.9 \, \mathrm{km}\,\mathrm{s}^{-1}$ \citep{Baumgardt_18}, being therefore consistent with the system's association to the cluster. This was already indicated by the close sky proximity to the centre of the cluster and the similarity of the DMs of all the other pulsars in the cluster \citep{Ridolfi_2022}.

%%%%%%%%%%%%%%%%%%%%%%%%%%%%%%%%%%%%%%%%%%%%%

\subsection{Keplerian orbital parameters}
\label{sec:keplerian}

With an orbital eccentricity of $e \, \simeq \, 0.89$, \psr\ is the third most eccentric known binary system in a GC, after NGC~6652A ($e \, \simeq \, 0.95$, \citealt{DeCesar_2015}) and Terzan 5 ap ($e \, \simeq \, 0.905$, \citealt{padmanabh_2024}). Using the values of the orbital period of the system and the {\it projected semi-major axis of the pulsar orbit} in time units ($x \equiv a_\mathrm{p}\sin i/c$) from Table \ref{table:timing_params}, we obtain for the mass function
%%%
\begin{equation}
f(\bMp, \bMc) 
= \frac{{({\bMc}\:{\sin{i}})}^{3}}{(\bMp + \bMc)^2} 
= \frac{4{\pi}^2}{{T_\odot}}\frac{{x}^3}{{{P_\mathrm{b}}^2}} 
= 0.1454280 (16) \,, \label{eq:mass_function}
\end{equation}
%%%
where the exact quantity $T_{\odot} \equiv ({\cal G M})^{\rm N}_{\odot}/c^3 = 4.925490947641266978\dots\, \upmu \rm s$ is the nominal solar mass parameter $({\cal G M})^{\rm N}_{\odot}$ (\citealt{PRSA2016}) in time units. 
In equations that contain $T_\odot$ explicitly, one has to use the adimensional mass parameters $\bar{M}_i \equiv GM_i/(\mathcal{GM})^{\rm N}_{\odot}$, ($i=\mathrm{p},\mathrm{c},\dots$), which are the numerical values of $M_i$ when expressed in units of solar mass ($\msun$). 

This high mass function value is indicative of a massive companion: assuming that the pulsar mass is larger than the smallest NS mass known ($\Mp\, =\, 1.17\, \msun$, \citealt{Martinez_2015}) and an edge-on orbit ($i = 90 \deg$), we derive $M_{\mathrm{c}} > 0.84\, \msun$. As discussed previously by \cite{Freire_2004, Freire_2007} and \citetalias{Ridolfi_19}, these combined factors reinforce the conclusion that \psr\ is the product of a secondary exchange encounter: the previous low-mass companion to the pulsar was replaced by its current massive counterpart.
The chaotic nature of such exchange encounters results in high orbital eccentricities. If the new companion to the pulsar is degenerate, then there is not much tidal circularisation.

\subsection{Rate of advance of periastron}
The updated timing model for the pulsar includes a very precise measurement of the rate of advance of periastron: $\dot\omega_{\rm{obs}}\:=\:0.012961(1)\: \rm{deg\: yr^{-1}}$. This is fully consistent with the value presented by \citetalias{Ridolfi_19} but twice as precise and is 40 times more precise than the first value quoted by \cite{Freire_2007}; this improvement is due to the extension of the dataset and the usage of pulse profile templates with consistent phase definition (as described earlier). This parameter is a quantification of the average change in the longitude of periastron of the system due to various relativistic and classical effects. 

Under the assumption that the $\dot\omega_\mathrm{obs}$ is purely relativistic, we can derive the total mass of the system from this measurement:
\begin{equation}
    \bMtot = \frac{1}{T_\odot}\left[{{\frac{\dot\omega_{\mathrm{Rel}}}{3}}{(1-e^2)}}\right]^{3/2}\left({\frac{P_{\mathrm{b}}}{2\pi}}\right)^{5/2} = 2.47346\,(28).
\end{equation}
This mass value is slightly larger and more than twice as precise than the last published work \citepalias{Ridolfi_19}; the stability of the observed $\dot\omega$ values (see Appendix~\ref{sec:PK_parameters_stability}) suggest that it is robust. 

However, there are additional effects that contribute to $\dot{\omega}_{\rm obs}$. The potentially largest is due to spin-orbit coupling ($\dot\omega_\mathrm{SO}$), if the companion WD is fast rotating. This includes the contributions due to Lense-Thirring precession and the one caused by the spin-induced quadrupole moment of the rotating WD companion \citep{Venkatraman_Krishnan_2020}. For the $1.09\,\msun$ companion of \psr\,(see next section), the maximum contribution, when the WD is rotating close to break-up, is $\dot\omega_\mathrm{SO} \approx 0.03 \,\dot\omega_\mathrm{obs}$. As will become clear below, even such an extreme rotation of the companion would have no influence on the conclusions of the present work.

The second largest is due to the interaction with a previously undetected third body, and is calculated in detail in Sect.~\ref{sec:external influence}. Additional contributions include $\dot\omega_{\rm{PM}}$, arising due to the proper motion of the system \citep{Kopeikin_1996}. For the measured proper motion of \psr\ as given in Table \ref{table:timing_params}, its maximum value is $\dot\omega_{\rm{PM}} = 8.72\,\times10^{-7}\,\mathrm{deg}\,\mathrm{yr}^{-1}$. This is comparable to the current measurement uncertainty in $\dot\omega_{\rm{obs}}$, and thus can be further neglected for all analysis.

The nominal measurement of $\dot\omega_{\rm{obs}}$ and its constraints are displayed in red in figure \ref{Mass_Mass_Diagram}. From the point where it intersects the constraint from the mass function (which excludes all points in gray), we derive a minimum companion mass of $\Mc = 0.96\,\msun$ and a maximum pulsar mass of $\Mp = 1.51\,\msun$. 

%%%%%%%%%%%%%%%%%%%%%%%%%%%%%%%%%%%%%%%%%%%%%
\subsection{Einstein Delay}
\begin{figure*}[!h]
    \centering
    \includegraphics[width=0.95\linewidth]{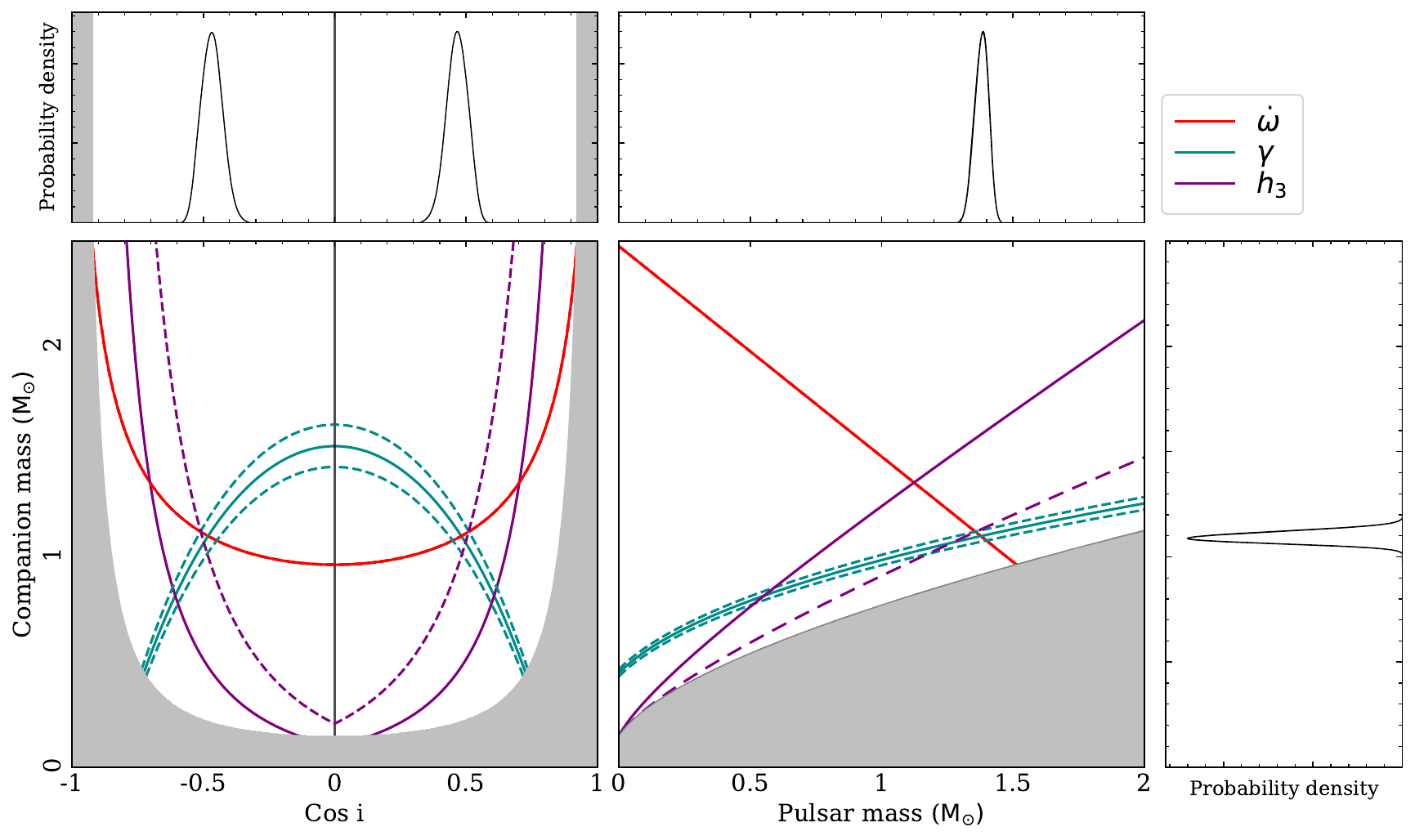}
    \caption{Orbital inclination and mass constraints for \psr. The solid and dotted lines represent the nominal values and 1-$\sigma$ constraints for the Post-Keplerian parameters (red: $\dot{\omega}$, cyan: $\gamma$, purple: $h_3$) obtained with the DDH Binary model. Bottom left plot: $\cos i$ versus the companion mass ($\Mc$); Bottom right plot: Pulsar mass ($\Mp$) versus $\Mc$.
    The grey portions on the two plots indicate the regions excluded for pulsar mass $(\Mp) \leq 0$ and $\sin{i} >1$ respectively. The normalised 1-D pdfs for $\cos i$, $\Mp$ and $\Mc$ 
    are shown in the top left, top right, and right panels.}
    \label{Mass_Mass_Diagram}
\end{figure*}

Owing to the long timing baseline of \psr, we have been able to precisely measure another relativistic parameter, the amplitude of the Einstein delay ($\gamma$) for the system. In GR, it is given by:
% \begin{flushright}
\begin{equation}
\label{eq:gamma}
\gamma  = {{T}^{2/3}_\odot}{e}
\left(\frac{P_{\mathrm{b}}}{2\pi}\right)^{1/3}{\frac{{\bMc}({\bMc + \bMtot})}{\bMtot^{4/3}}}.
\end{equation}
% \end{flushright}
This delay quantified by $\gamma$ results from the combination of the varying special relativistic time dilation due to the changing velocity of the pulsar in its orbit and the varying gravitational redshift due to the companion. The Einstein delay was earlier measured by \citetalias{Ridolfi_19}, and this allowed for the determination of the individual masses for the system. However, in this work, we report an updated value, $\gamma = 0.01925\,(48)\,\mathrm{s}$.  The constraints on $\gamma$ are indicated by the cyan lines in Fig.~\ref{Mass_Mass_Diagram}. This value is 1.8 times more precise than the value determined by \citetalias{Ridolfi_19}, but it is also $2.6\,\sigma$ smaller. This difference is discussed in more detail in Appendix~\ref{sec:PK_parameters_stability}.

With this measurement and the knowledge of $\mtot$ as derived above, we derive the masses using equations (21) and (22) from \citetalias{Ridolfi_19}. For \psr, under the assumption of GR, we obtain $\Mc = 1.09\,(2)\,\msun$, $\Mp =  1.38\,(2)\,\msun$ and $i = 61.5 \deg$ or $i = 118.5 \deg$. These differ from the mass estimates of \citetalias{Ridolfi_19} because of the difference in the measured $\gamma$.

As discussed by \citetalias{Ridolfi_19}, for wide binaries the measurement of $\gamma$ is covariant with secular variations of the projected semi-major axis ($\dot{x}$). In addition, they assumed that the $\dot{x}$ of \psr\, originates from the proper motion of the system; this is an effect that must be taken into account for any wide binary system. However, the realization that the \psr\, system has a nearby companion means that perturbations from that third object can also contribute to $\dot{x}$ ($\dot{x}_\mathrm{tidal}$), and therefore affect $\gamma$ and the mass measurements. The magnitude of this effect will be estimated in Sect.~\ref{sec:external influence}, and the individual masses will be estimated in Sect.~\ref{sec:chi2map}.

%%%%%%%%%%%%%%%%%%%%%%%%%%%%%%%%%%%%%%%%%%%%%%%%%%%%%%%%%%%%%%%%%%%%%%%%%%%%%%%%%%%%%%%%%%
%% Mass-Mass Diagram ---------------------------
%%%%%%%%%%%%%%%%%%%%%%%%%%%%%%%%%%%%%%%%%%%%%%%%%%%%%%%%%%%%%%%%%%%%%%%%%%%%%%%%%%%%%%%%%%

\subsection{Shapiro delay}\label{ShapiroDelay}

With an aim to quantify the relativistic light-propagation delay in the system, we have taken extensive amounts of data at periastron and superior conjunction. In our DDH orbital model, we fixed an estimate of the orthometric ratio $\stig$ derived from $i$, $\varsigma = \frac{\sin{i}}{1 + |\cos{i}|} = 0.59$; from this we obtained a measurement of the orthometric amplitude $h_3 = 0.52\,(53)\,\upmu\mathrm{s}$. This indicates that we do not detect any Shapiro delay for the system. This is in agreement with the expected non-significant measurement due to the face-on inclination of the system derived from the other PK parameters.

\section{Period derivatives of a pulsar in a globular cluster} \label{sec:period derivatives}
% \section{Period derivatives: A theoretical interpretation} \label{Theo}

Many of the new parameters we measure in this work are the derivatives of the spin and orbital frequencies. Given the common origin of these terms, and some of the subtleties in their analysis, we discuss them at length in this dedicated section. This includes new material that is relevant for the analysis of, in principle, all pulsars in GCs where statistically significant changes in the acceleration are measurable, i.e., the vast majority of GC pulsars with long timing baselines. In the following section, we will discuss the physical causes of the acceleration of this system and its variation with time.

\subsection{A theoretical interpretation} \label{subsec:Theo}

As we cannot measure radial motion, we observe modified ``Doppler-shifted'' values of the intrinsic spin period of the pulsar ($P_\mathrm{int}$) and orbital period of the binary ($P_\mathrm{b,int}$). The observed parameters are related to these values as: 
\begin{subequations}
\begin{align}
    & {P}_{\mathrm{obs}} = D^{-1}\,P_{\mathrm{int}} \simeq {P}_{\mathrm{int}} \left[1 + {{V}_\mathrm{l}}/{c}\,\right], \label{Obs_period} \\
    & {P}_{\mathrm{b,obs}} = D^{-1}\,P_{\mathrm{b,int}} \simeq {P}_{\mathrm{b,int}} \left[1 + {{V}_\mathrm{l}}/{c}\,\right], 
    \label{Obs_OrbPer}
\end{align}
\end{subequations}
where $D$ is the Doppler shift factor \citep{Damour_Taylor_1992}, the last term is a leading order term and ${V}_\mathrm{l}$ is given by
\begin{equation}
V_\mathrm{l} = \vec{V}_\mathrm{CM} \cdot \hat{\vec{K}}_0,
\end{equation}
where $\vec{V}_\mathrm{CM}$ is the velocity of the centre of mass of the binary (CM) relative to the Solar System barycentre (SSB) and $\hat{\vec{K}}_0$ is the unit vector pointing from the SSB to CM. Although this radial motion also affects other distance and mass parameters of the binary, these contributions are negligible compared to the observed values and hence ignored in further analysis. 

A change of $V_\mathrm{l}$ with time, given by $a_\mathrm{l} = \mathrm{d} V_\mathrm{l} / \mathrm{d} t$, causes a secular change in the Doppler shift. Differentiating equations~\ref{Obs_period} and ~\ref{Obs_OrbPer}, we obtain \citep{Phinney_1993}:
\begin{subequations}
\begin{align}
     & \left(\frac{\dot{P}}{P}\right)_{\mathrm{obs}} = \left(\frac{\dot{P}}{P}\right)_{\mathrm{int}} + \frac{a_\mathrm{l}}{c}, \label{Pdot}\\
     & \left(\frac{\dot{P_\mathrm{b}}}{P_\mathrm{b}}\right)_{\mathrm{obs}} = \left(\frac{\dot{P_\mathrm{b}}}{P_\mathrm{b}}\right)_{\mathrm{int}} + \frac{a_\mathrm{l}}{c}.  \label{Pbdot}
\end{align}
\end{subequations} 
For a pulsar in a GC, $a_{\rm l}$ has contributions from multiple factors, and can be explicitly written as
\begin{equation}
    a_{\rm l} = a_{\rm{Shk}} + {a_{\rm{Gal}}} + {a_{\rm{GC}}} + {a_{\rm{NS}}}, \label{eqn:acceleration contributions}
\end{equation}
where $a_{\rm{Shk}}$ indicates the apparent acceleration due to Shklovskii effect \citep{Shklovskii_1970}, $a_{\rm{Gal}}$ is the difference between the accelerations of the Solar System and NGC~1851 in the field of the Galaxy,
$a_{\rm{GC}}$ is the line-of-sight acceleration of the pulsar in the gravitational field of the GC and 
$a_{\rm{NS}}$ is the acceleration in the field of nearby stars.
A detailed investigation of the various accelerations is presented in Sect.~\ref{sec:external influence}. 

To derive the second order spin period and orbital period derivatives of the system, we differentiate equations~\ref{Pdot} and ~\ref{Pbdot}. For a pulsar in a globular cluster, the change in the line-of-sight acceleration is much larger than the change in the first term on the right of these equations and thus dominates the contribution to the second order derivatives \citep{Phinney_1992}. Under this consideration, and using equation (7) from \citet{Freire_2017}, we have 
\begin{eqnarray}
\begin{aligned}
     & \left(\frac{\ddot{P}}{P}\right)_{\mathrm{obs}} \simeq \frac{\dot{a}_\mathrm{l}}{c}  \simeq -\left(\frac{\ddot{f}}{f}\right), \label{Pddot}\\
    & \left(\frac{\ddot{P_\mathrm{b}}}{P_\mathrm{b}}\right)_{\mathrm{obs}} = \frac{\dot{a}_\mathrm{l}}{c}  \simeq -\left(\frac{\ddot{f_\mathrm{b}}}{f_\mathrm{b}}\right), \label{Pbddot}
\end{aligned}
\end{eqnarray} 
where $f$ and $f_\mathrm{b}$ denotes the spin frequency of the pulsar and the orbital frequency of the binary, respectively. Note that in these equations, any ``intrinsic'' terms are very likely to be negligible, a point of importance for what follows.

Equations~\ref{Pddot} can be differentiated to get quite accurate expressions for further higher order derivatives, for $k>2$, as
\begin{eqnarray}
\begin{aligned}
     & \frac{1}{P}\frac{d^{k}{P}}{dt^{k}} = \frac{1}{c}\frac{d^{k-1}{a}_\mathrm{l}}{dt^{k-1}} \simeq \frac{f^{(n)}}{f}, \label{Pdot_n}\\
    & \frac{1}{P_\mathrm{b}}\frac{d^{k}{P_\mathrm{b}}}{dt^{k}} = \frac{1}{c}\frac{d^{k-1}{a}_\mathrm{l}}{dt^{k-1}} \simeq \frac{f_\mathrm{b}^{(n)}}{f_\mathrm{b}}. \label{Pbdot_n}\\
\end{aligned}
\end{eqnarray}

The values of all the spin and orbital period derivatives for \psr\ are given in Table~\ref{table:evolution_derivatives}, and a detailed discussion for each of them is given below. 

%%%%%%%%%%%%%%%%%%%%%%%%%%%%%%%%%%%%%%%%%%%%%%%%%%%%%%%%%%%%%%%%%%%%%%%%%%%%%%%%%%%%%%%%%%%%%%%%%%%%%%%%%%%%%%%%%%%%%%%%%%%%%%%%%%%%%%%%%%%%
%%%%%%%%%%%%%%%%%%%%%%%%%%%%%%%%%%%%%%%%%%%%%%%%%%%%%%%%%%%%%%%%%%%%%

\begin{table*}[h!]
%\large
\caption{Spin and orbital frequency derivatives for \psr\ for two different reference epochs.
The parameters are derived from a modified DDH timing solution,
except a) the higher order orbital frequency derivatives, which are derived from the spin frequency derivatives (names in italic) and b) the derived parameters below the horizontal lines. The values in the parenthesis indicate 1-$\sigma$ uncertainties on the last digit.}
\centering 
% \begin{scriptsize}
\begin{tabular}{lccc}
\hline\hline                 % inserts double horizontal lines
Reference epoch (MJD) & 53623.155 (Epoch 1) &  53623.155 (Epoch 1) & 59728.338 (Epoch 2) \\ 
Epoch of periastron, $T_{0}$ (MJD) & 53623.15508797(35) & 53623.15508755(33) & 59728.33832689(13) \\
 & \citepalias{Ridolfi_19} & (This work) & (This work) \\
\hline
\multicolumn{3}{c}{With one orbital period derivative} \\
\hline\hline
   Spin frequency, $f$ (Hz) & 200.37770740535 (10) & 200.3777074057 (2) & 200.37770716645 (1) \\
   First spin frequency derivative, $\dot{f}$ ($10^{-17} \mathrm{Hz}\,\mathrm{s}^{-1}$)& $-$2.8 (5) & $-$3.5 (4)  & $-$94.51 (8) \\ 
   Second spin frequency derivative, $\ddot{f}$ ($10^{-24} \mathrm{Hz}\,\mathrm{s}^{-2}$)& $-$1.533 (27) & $-$1.31 (2) & $-$2.14 (1) \\ 
   Third spin frequency derivative, $\dddot{f}$ ($10^{-33} \mathrm{Hz}\,\mathrm{s}^{-3}$)& - & $-$1.57 (8) & $-$1.57 (8) \\
   Orbital period, $P_\mathrm{b}$ (day) & 18.785179217(19) & 18.785179201 (2) & 18.785179194 (2) \\
   Observed orbital period derivative, $\dot{P}_\mathrm{b,obs}$ ($10^{-12}$ s s$^{-1}$) & 22 (9) & $-$1.2 (7) & $-$1.2 (7)\\
   \midrule 
   Intrinsic spin period derivative, $\dot{P}_\mathrm{int}$ ($10^{-21}\, \mathrm{s}\,\mathrm{s}^{-1}$) & - & 4.05 (9) & 26.71 (2)\\
   Characteristic age (Gyr) & - & 19.5 (4) & 2.960 (2) \\
   \hline
    \multicolumn{3}{c}{With multiple orbital period derivatives}\\
    \hline\hline
   Spin frequency, $f$ (Hz) & - & 200.3777074057 (2) & 200.37770716645 (1) \\
   First spin frequency derivative, $\dot{f}$ ($10^{-17} \mathrm{Hz}\,\mathrm{s}^{-1}$) & - & $-$3.5 (4) & $-$94.51 (8)\\ 
   Second spin frequency derivative, $\ddot{f}$ ($10^{-24} \mathrm{Hz}\,\mathrm{s}^{-2}$)& - & $-$1.31 (2) & $-$2.14 (2)\\ 
   Third spin frequency derivative, $\dddot{f}$ ($10^{-33} \mathrm{Hz}\,\mathrm{s}^{-3}$)& - & $-$1.57 (8) & $-$1.57 (8) \\
   {\it Second orbital frequency derivative}, $\ddot{f}_\mathrm{b}$  ($10^{-33}$ Hz s$^{-2}$) & - & $-$4.03 & $-$6.58\\
   {\it Third orbital frequency derivative},$\dddot{f}_\mathrm{b}$ ($10^{-42}$ Hz s$^{-3}$) & - & $-$4.84 & $-$4.84 \\
   Orbital period, $P_\mathrm{b}$ (day) & - & 18.785179207 (2) & 18.785179196 (2) \\
   Observed orbital period derivative, $\dot{P}_\mathrm{b,obs}$ ($10^{-12}$ s s$^{-1}$) & - &  $-$5.2 (7) & 2.2 (7)\\
   \midrule 
   Intrinsic spin period derivative, $\dot{P}_\mathrm{int}$ ($10^{-20}\, \mathrm{s}\,\mathrm{s}^{-1}$) & - & 1.633 (9) & 1.632 (2)\\
   Characteristic age (Gyr) & - & 4.84 (3)  & 4.844 (6) \\
\hline\hline                              
\label{table:evolution_derivatives}
\end{tabular}
\end{table*}

%%%%%%%%%%%%%%%%%%%%%%%%%%%%%%%%%%%%%%%%%%%%%%%%%%%%%%%%%%%%%%%%%%%%%%%%%%%%%
%%%%%%%%%%%%%%%%%%%%%%%%%%%%%%%%%%%%%%%%%%%%%%%%%%%%%%%%%%%%%%%%%%%%%%%%%%%%%
%%%%%% 

%%%%%%%%%%%%%%%%%%%%%%%%%%%%%%%%%%%%%%%%%%%%%%%%%%%%%%%%%%%%%%%%%%%%%%%%%%%%%
%%%%%%%%%%%%%%%%%%%%%%%%%%%%%%%%%%%%%%%%%%%%%%%%%%%%%%%%%%%%%%%%%%%%%%%%%%%%%
%%%%%% 
%%%%%%%%%%%%%%%%%%%%%%%%%%%%%%%%%%%%%%%%%%%%%%%%%%%%%%%%%%%%%%%%%%%%%%%%%%%%%
%%%%%%%%%%%%%%%%%%%%%%%%%%%%%%%%%%%%%%%%%%%%%%%%%%%%%%%%%%%%%%%%%%%%%%%%%%%%%

\subsection{Measured spin period (frequency) derivatives} \label{subsec:spin frequency derivatives}
The period derivatives of pulsars in GCs, unlike the ones in the Galactic disk, are highly influenced by the gravitational field of the globular cluster and nearby stars. In this section, we discuss the spin period (equivalently, spin frequency) derivatives for \psr. These are measured for two different reference epochs: Epoch 1 (MJD = 53623, 10 September 2005)
was used by \cite{Freire_2007} and \citetalias{Ridolfi_19}, Epoch 2 (MJD = 59728, 29 May 2022) is used in Table~\ref{table:timing_params}. This latter epoch is within the recent phase of intense MeerKAT timing; for this reason the timing parameters in Table~\ref{table:timing_params} have smaller correlations than if they were measured for Epoch 1.

\subsubsection{First spin frequency derivative} \label{subsubsec:FSFD}

We report an updated and much more precise measurement of the first spin frequency derivative for \psr\ (see Table~\ref{table:evolution_derivatives}), which translates into a spin period derivative of $\dot{P}_{\mathrm{obs}} = 2.3538\,(20)\,\times\,10^{-20}\,{\mathrm{s}}\,{\mathrm{s}^{-1}}$. This is $\sim$$34$ times larger than the value reported by \citetalias{Ridolfi_19}, but measured for Epoch 2. This change is predominantly due to a large variation in the acceleration of the system: if we instead use Epoch 1, we obtain a value of $\dot{f}$ that is consistent with that presented in those works, but still $\sim 27$ times smaller than the value measured for Epoch 2. We now discuss this large change of $\dot{P}$.

%%%%%%%%%%%%%%%%%%%%%%%%%%%%%%%%%%%%%%%%%%%%%%%%%%%%%%%%%%%%%%%%%%%%%
%%%%%%%%%%%%%%%%%%%%%%%%%%%%%%%%%%%%%%%%%%%%%%%%%%%%%%%%%%%%%%%%%%%%%

\subsubsection{Higher spin frequency derivatives} \label{subsubsec: higher spin frequency derivative}

In the extremely dense core of a GC, a pulsar experiences variations of its acceleration due to its movement in the gravitational potential of the cluster and its interaction with the nearby stars \citep{Phinney_1992}. Equation~\ref{Pddot} shows that this change in the line-of-sight acceleration of the system, also known as the line-of-sight `jerk' ($\dot{a}_\mathrm{l}$), is manifested by the second period (frequency) derivative of the pulsar (Table~\ref{table:evolution_derivatives}). From its value at Epoch 2, we estimate the jerk for \psr\, as
\begin{equation}
    {\dot{a}_{l}} = 3.20\,\times\,10^{-18}\,{\mathrm{m}}\,{\mathrm{s}^{-3}} \label{1851A_jerk}.
\end{equation}
Although a similar value for $\ddot{f}$ was measured by \citetalias{Ridolfi_19}, they did not comment on its unusual size, which is orders of magnitude larger than the observed values for other MSPs in globular clusters \citep{Freire_2017,Prager_2017}, and only a factor of 5 smaller than observed in PSR~B1620$-$26, the triple system in the globular cluster M4 \citep{TACL_1999}. This is so large that it causes the large variation of $\dot{P}_\mathrm{obs}$ between Epochs 1 and 2. Inevitably, this jerk affects the observed orbital period and its derivatives, as discussed in Sect.~\ref{subsec: orbital period derivatives}. As mentioned above, we discuss the implications of this unusually large jerk in Sect.~\ref{sec:external influence}. 

The long timing baseline and the precise timing solution allowed for a $19.4\,\sigma$ measurement of the third spin frequency derivative for \psr\, as given in Table~\ref{table:evolution_derivatives}. Its negative value implies that the large jerk (with a negative sign in $\ddot{f}$) is becoming even larger in magnitude. For a pulsar in the core of such a dense GC, such higher order derivatives of the spin period are fairly common: unlike the lower spin frequency derivatives, which can be caused by the mean field of the GC, the higher order derivatives are caused by the gravitational fields of nearby stars \citep{Phinney_1992}. 

%%%%%%%%%%%%%%%%%%%%%%%%%%%%%%%%%%%%%%%%%%%%%%%%%%%%%%%%%%%%%%%%%%%%%
%%%%%%%%%%%%%%%%%%%%%%%%%%%%%%%%%%%%%%%%%%%%%%%%%%%%%%%%%%%%%%%%%%%%

\subsubsection{Evolution of spin frequency derivatives} \label{subsubsec:Evolution_SFD}

\begin{figure}[hbt!]
    \centering
    \includegraphics[width=\hsize, height =10cm]{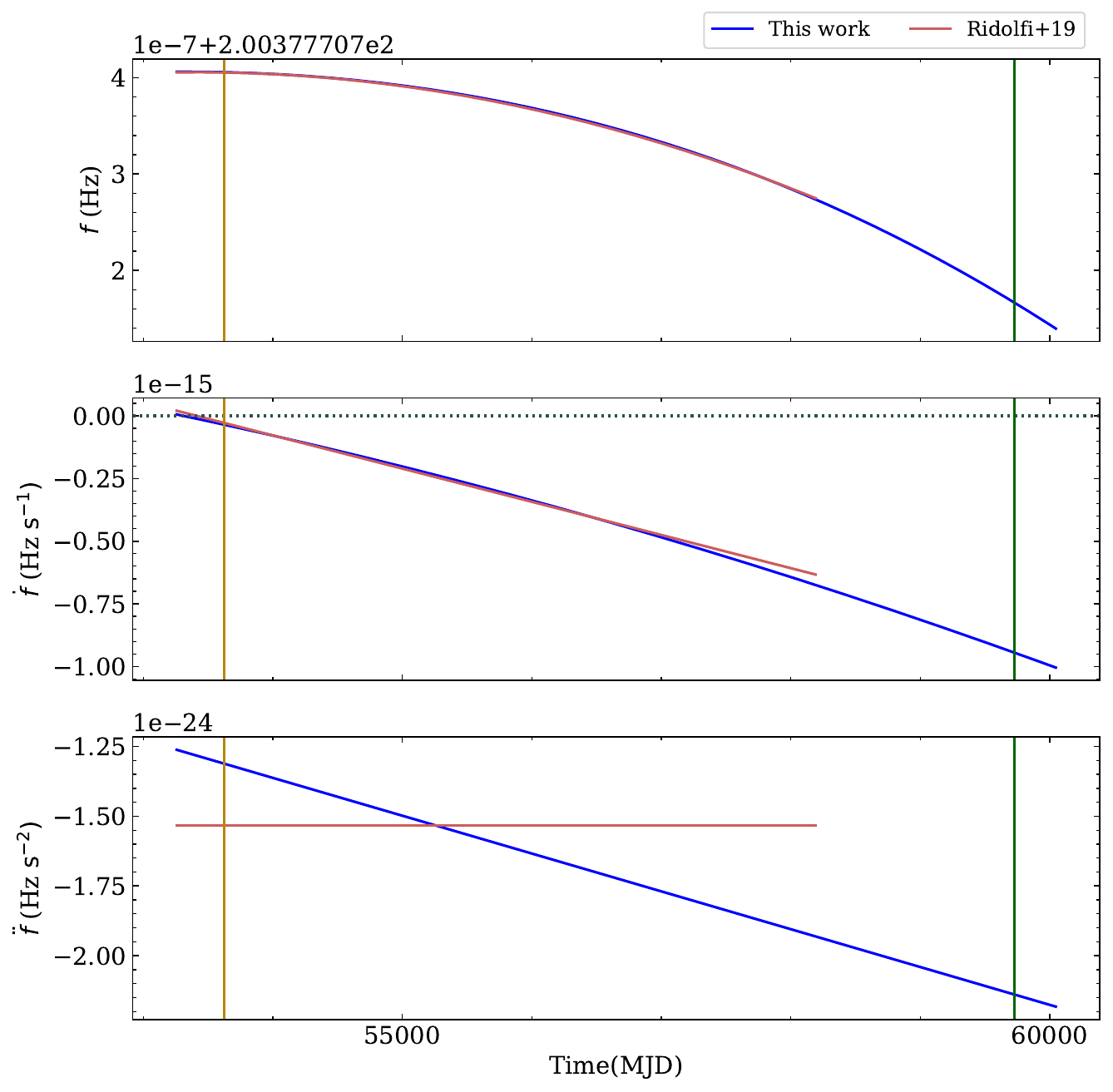}
    \caption{\textit{Top to bottom}: Evolution of the observed spin frequency ($f$), its first derivative ($\dot{f}$), and its second derivative ($\ddot{f}$) as a function of time (MJD). The vertical yellow and green lines indicate Epochs 1 and 2 respectively. The values from \citetalias{Ridolfi_19} are shown in red, and the results from this work are shown in blue. The large change in the $\dot{f}$ value between the two epochs can be clearly seen.}
    \label{RD_PeriodDerivatives}
\end{figure}

We now look in detail at the evolution of the spin frequency ($f$) and its time derivatives ($\dot{f}$, $\ddot{f}$, and $\dddot{f}$), using them as the coefficients in a Taylor expansion \citep{Lorimer_Kramer}, 
\begin{equation}
    f(t) = f_{0} + \dot{f}_{0}(t-t_0) + \frac{1}{2}\ddot{f}_{0}(t-t_0)^{2} + \frac{1}{6}{\dddot{f}_{0}}(t-t_0)^{3} + ... , \label{eqn: evolution of derivatives}
\end{equation}
around Epoch 1 and 2. Figure \ref{RD_PeriodDerivatives} shows the evolution of the spin frequency and its first and second derivatives over the total timing baseline of this pulsar. As discussed previously, the high jerk for the system is clearly evident from the evolution of the first spin frequency derivative with time. In the last plot, we can also appreciate how fast the jerk is increasing with time, having increased by about 65\% between Epochs 1 and 2.

Despite these changes between Epochs 1 and 2, and between our Epoch 1 and the values measured by \citetalias{Ridolfi_19}, we can see that the spin evolution we observe is mostly consistent with that described by them, at least for the period of time covered by their earlier timing solution.

%%%%%%%%%%%%%%%%%%%%%%%%%%%%%%%%%%%%%%%%%%%%%%%%%%%%%%%%%%%%%%%%%%%%%
%%%%%%%%%%%%%%%%%%%%%%%%%%%%%%%%%%%%%%%%%%%%%%%%%%%%%%%%%%%%%%%%%%%%
%%%%%%%%%%%%%%%%%%%%%%%%%%%%%%%%%%%%%%%%%%%%%%%%%%%%%%%%%%%%%%%%%%%%%
%%%%%%%%%%%%%%%%%%%%%%%%%%%%%%%%%%%%%%%%%%%%%%%%%%%%%%%%%%%%%%%%%%%%
%%%%%%%%%%%%%%%%%%%%%%%%%%%%%%%%%%%%%%%%%%%%%%%%%%%%%%%%%%%%%%%%%%%%%
%%%%%%%%%%%%%%%%%%%%%%%%%%%%%%%%%%%%%%%%%%%%%%%%%%%%%%%%%%%%%%%%%%%%

\subsection{Orbital period derivatives} \label{subsec: orbital period derivatives}

The extension of the timing baseline was hugely beneficial for a refined measurement of the change in the orbital period for \psr. The inclusion of orbital frequency derivatives higher than first order is, as explained below, necessary for a correct description of the orbital evolution of the system, and has far-reaching implications. 

\subsubsection{First orbital period derivative} \label{subsubsec: first orbital period derivative}

Another quantity that has changed considerably relative to \citetalias{Ridolfi_19} is the measurement of the orbital period derivative of the system. Equation~\ref{Obs_OrbPer} shows that the observed change in the orbital period, like the spin period, is also affected significantly by a secular change in the Doppler shift; thus the acceleration should affect the $\dot{P}$ and $\dot{P}_\mathrm{b}$ of a pulsar in a globular cluster in a similar way. The unexpectedly large positive value of $\dot{P}_{\rm b}$ observed by \citetalias{Ridolfi_19} could not be explained by a large acceleration contribution from the cluster's field: such an acceleration would result in a value of $\dot{P}$ that is much larger than that observed. The expected intrinsic variation in the orbital period ($\dot{P}_{\rm b, int}$) from orbital decay due to gravitational wave emission is orders of magnitude smaller than the observed value, and also of negative sign, and thus could not account for the large positive value of $\dot{P}_{\rm b}$. 

In this work, we have revisited this problem with a much longer dataset and planned observations to check the robustness of the measurement. The updated value of the $\dot{P}_{\rm b, obs}$ for this system, derived using the DDH model, is given in Table~\ref{table:evolution_derivatives}. This measurement is 13 times more precise than the estimate by \citetalias{Ridolfi_19}, and 2.6-$\sigma$ smaller.
The reasons for the difference are in part related to the discovery of the 2-second clock offset between the earlier GMRT dataset, but also to the decrease in correlation between different parameters. Furthermore, as we'll see below, the jerk should also cause this value to change with time.

Knowing the masses (from $\dot{\omega}$ and $\gamma$), we can use GR (using the equation of \citealt{Peters_1964}) to estimate the orbital decay due to the emission of gravitational waves, which we assumed is the intrinsic variation of the orbital period. This gives $\dot{P}_{\rm b, int} = -0.153\times10^{-12}\,\mathrm{s}\, \mathrm{s}^{-1}$, which is one order of magnitude smaller than $\dot{P}_{\rm b, obs}$; this means that the latter is dominated by the contribution from $a_l$.

\subsubsection{Higher orbital frequency derivatives} \label{subsubsec: higher orbital freq derivatives}

Higher order orbital frequency derivatives are generally not measurable for MSPs in globular clusters, unless they are in ``spider'' systems \citep{Freire_2017}; in these cases the variations are not related to variations in the accelerations of those systems and arise due to the changing gravitational quadrupole moment of
their companion (e.g. \citealt{Ridolfi_2016}). The large line-of-sight jerk for \psr\ , which changes the $\dot{P}_\mathrm{obs}$ by a factor of 27 between the two reference epochs, should cause a similar correlated change in the $\dot{P}_\mathrm{b,obs}$ for the pulsar. Even though the latter effect is not clearly measurable, it should be taken into consideration; as we'll see below this makes a real difference.

A precise description of the variation of the orbital phase requires the inclusion of higher order derivatives in the orbital frequency ($f_\mathrm{b} = 1/P_\mathrm{b}$), described as orbital frequency derivatives ($f_\mathrm{b}^{(n)}$) for the system. Since the higher derivatives of $f$ and $f_\mathrm{b}$ are affected solely by the derivatives of the acceleration -- as mentioned above, there should be no intrinsic contributions to them -- the $f_\mathrm{b}^{(n)}$ $(n \, \geq\,2)$ values can be estimated directly from the corresponding values of $f^{(n)}$ by re-writing eqs.~\ref{Pbdot_n}:
\begin{equation}
f_\mathrm{b}^{(n)} = f^{(n)} \frac{f_\mathrm{b}}{f},
\label{eq:dfbn}
\end{equation}
which assumes that these derivatives are measured at the same time.
Such an estimate is intrinsically much more precise than any direct measurement of $f_\mathrm{b}^{(n)}$ given the much higher precision in the measurement of $f^{(n)}$, i.e., in the measurement of the spin phase relative to the orbital phase. We have modified our timing models to take into account the jerk and its variation in a self-consistent way using these equations (see Appendix~\ref{sec:DD_modification}). The values of $f_\mathrm{b}^{(n)}$ derived from the values of $f^{(n)}$ are quoted in the second part of Table~\ref{table:timing_params}.

As we will see next, a direct argument for the necessity of taking the $f_\mathrm{b}^{(n)}$ into account comes from the inconsistent estimates of intrinsic spin period derivative, $\dot{P}_{\mathrm{int}}$ (and therefore of quantities derived from it assuming rotating dipole model, like characteristic age, surface magnetic field and spin-down power) for Epochs 1 and 2, that we obtain when the change in the orbital period is estimated by a single derivative. The agreement of the values of $\dot{P}_{\mathrm{int}}$ after the addition of $\ddot{f}_\mathrm{b}$ and $\dddot{f}_\mathrm{b}$ can be seen in the second part of Table~\ref{table:evolution_derivatives}. 

\subsection{Intrinsic spin period derivative and the characteristic age} \label{subsec: intrinsic spin period derivative}

The similarity of equations~(\ref{Obs_period}) and (\ref{Obs_OrbPer}) means that the difference of these two can be used to estimate the intrinsic change in the spin period of the pulsar as \begin{equation}
    \dot{P}_{\mathrm{int}} =  {P}\left(\frac{\dot{P}_{\mathrm{b,int}} - \dot{P}_{\mathrm{b,obs}}} {{P_{\mathrm{b}}}}\right) + \dot{P}_{\mathrm{obs}}. 
    \label{Pdot_int}
\end{equation}
In Table~\ref{table:evolution_derivatives}, we calculate this parameter and the characteristic age for Epochs 1 and 2, using the equation in \cite{Lorimer_Kramer}, from the orbital and spin period derivatives measured at each epoch.
In the first part of the table, we see how neglecting any higher orbital period derivatives produced completely different `local estimates' of $\dot{P}_\mathrm{int}$. 
In the second part of the table, we see how taking the higher order estimates of  $\ddot{f}_\mathrm{b}$ into account
leads to consistent estimates for $\dot{P}_{\mathrm{int}}$. This quantity is intrinsic to the pulsar and (for MSPs) should not change significantly with the time at which it is estimated. This highlights the importance of taking these higher order orbital frequency derivatives into account.

From $\dot{P}_{\mathrm{int}}$, we estimate a characteristic age $\tau_\mathrm{c}=4.8\,\mathrm{Gyr}$; for the magnetic field at the surface we obtain $B = 2.8\,\times10^{8}\,{\mathrm{G}}$, making this pulsar a rather typical MSP.

%%%%%%%%%%%%%%%%%%%%%%%%%%%%%%%%%%%%%%%%%%%%%%%%%%%%%%%%%%%%%%%%%%%%%%%%%%%%%
%%%%%%%%%%%%%%%%%%%%%%%%%%%%%%%%%%%%%%%%%%%%%%%%%%%%%%%%%%%%%%%%%%%%%%%%%%%%%

%%%%%%%%%%%%%%%%%%%%%%%%%%%%%%%%%%%%%%%%%%%%%%%%%%%%%%%%%%%%%%%%%%%%%%%%%%%%%%%%

\section{Investigating the external influence on the binary}
\label{sec:external influence}

\subsection{On the need for a third object}

It is quite evident from Fig.~\ref{RD_PeriodDerivatives} that the pulsar's apparent spin-up changed to an apparent spin-down just a few years before Epoch 1: this, and the continued large change in the apparent spin-down between Epochs 1 and 2 hint towards the fact that the system is being substantially perturbed in the globular cluster environment.

As pulsars slow down due to the loss of rotational energy, the intrinsic spin period derivative (first term on the right of equation~\ref{Pdot}) has a small, positive and constant value. The measured value of $\dot{P}_\mathrm{obs}$ thus implies an upper limit on the line-of-sight acceleration of the pulsar, and the latter estimated at Epoch 2 using equation~\ref{Pdot}, is
${a_\mathrm{l,max}} = c ({\dot{P}/P})_{\rm obs} = 1.41\,\times\,10^{-9}\,{\mathrm{m}}\,{\mathrm{s}^{-2}} $. 

The various contributions to the acceleration are given explicitly in equation~\ref{eqn:acceleration contributions}, and the values are discussed here. In this work, we have been able to precisely measure the total proper motion of the system, $\mu_\mathrm{T} = 2.76\,(12)$ \masy.
This, and the distance to the cluster ($d \approx 11.7$ kpc, \citealt{Libralato_2022}) yield $a_{\mathrm{Shk}} = {\mu^2} d = 6.5\, (6)\, \times 10^{-11}$ m s$^{-2}$. Using the Milky Way potential from \cite{McMillan2017} and the Galactic coordinates of the globular cluster, we obtain $a_{\mathrm{Gal}} = -1.174\, \times 10^{-11}\,{\mathrm{m}}\,{\mathrm{s}^{-2}}$.
The sum of $a_{\mathrm{Shk}}$ and $a_{\mathrm{Gal}}$ is about 27 times smaller than ${a_\mathrm{l,max}}$ at Epoch 2 given above. For most pulsars, the explanation for this is that $a_{\rm{GC}}$ is much larger, and the effect of the passing stars ($a_{\rm{NS}}$) is negligible in comparison \citep{Phinney_1993}. However, the large change in the line of sight acceleration between Epochs 1 and 2 suggests otherwise. 

The $a_{\rm{GC}}$ term can also change with time. NGC~1851 has a high velocity dispersion of 10.2 km s$^{-1}$ and a small core radius of 5\farcs4 \citep{Miocchi2013}; using these values and equation (10) from \cite{Freire_2017}, we estimate the maximum line-of-sight jerk that could be induced by the gravitational potential of the cluster. We use the escape velocity of the cluster, 42.9 km s$^{-1}$ \citep{Baumgardt_18}, as the maximum velocity of the pulsar relative to the cluster, and the projected separation of \psr\ from the core of the cluster as given in Table~\ref{table:timing_params}. We derive 
\begin{equation}
    |\dot{a}_\mathrm{l,GC,max}| = 1.40\, \times\, 10^{-19}\, \rm{m}\, \mathrm{s}^{-3} \label{max_GC_jerk},
\end{equation}
which is 23 times smaller than the observed jerk for the system. This suggests that $\dot{a}_\mathrm{l}$ is not due to the mean field of the cluster, but from the varying accelerations due to close encounters with nearby stars, which can be significantly larger \citep{Phinney_1993, Prager_2017}. 

Not only does this make it evident that the most likely cause for the large observed jerk is an external mass, but it also opens up two scenarios. If the third mass is bound to the inner binary, we have a system similar to PSR~B1620$-$26  \citep{Thorsett_1993, TACL_1999}; such triple systems could be formed through dynamical encounters in these dense environments. In the alternative scenario, the third body is not bound to the inner binary and is merely flying by.
Such a phenomenon has not been previously observed for any pulsar in a globular cluster. The probability of this for the M4 triple system was negligible \citep{TACL_1999}; however, given the extremely dense and compact core of NGC~1851, this possibility cannot be ruled out for \psr. 

We determined the closest distance of approach ($b_\mathrm{star}$) of a nearby mass to induce the observed jerk for \psr\ (equation 4.4 from \citealt{Phinney_1993}), and the probability of a passing star being so close is given by \citep{Phinney_1993}
\begin{equation}
p_\mathrm{star}(<b_{\rm star}) = 1 - \mathrm{exp}\left[-(4\pi/3)\,{n}\,{b^{3}_{\rm star}}\right],
\end{equation}
where $n$ is the central density of the globular cluster. For a star of mass 1 $\msun$, the estimated closest approach is 498 astronomical units (AU), and the corresponding probability is 0.196. The variation of the $b_\mathrm{star}$ and $p_\mathrm{star}$ values as a function of masses are shown in Fig.~\ref{Mass_bValues} below. These values suggest that for \psr, the probability of a flyby is a fair possibility, owing to the high stellar density of NGC 1851. In the next section, we discuss further implications for these values and investigate the nature of the external influence in bound and unbound orbits.

\begin{figure}[hbt!]
\centering
    \includegraphics[width=\columnwidth]{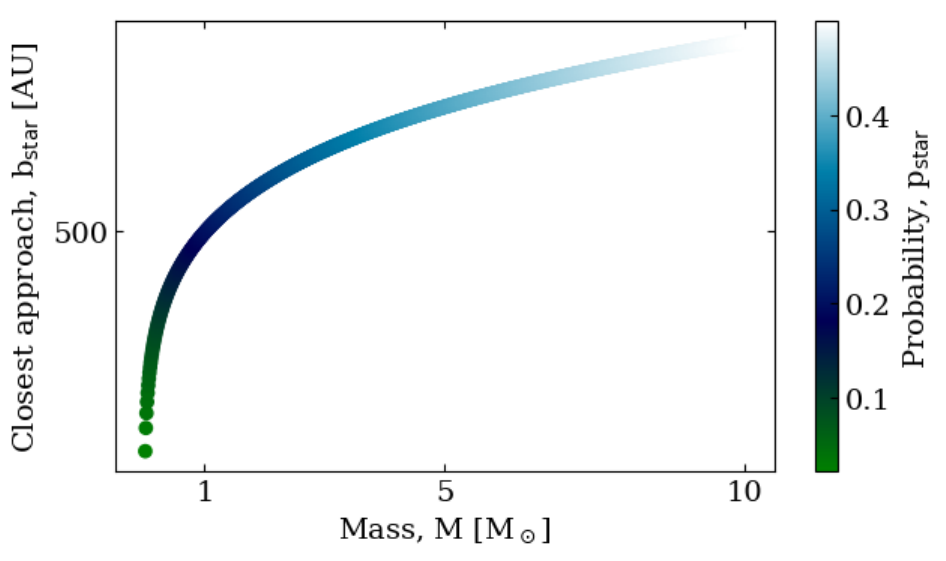}
    \caption{The closest distance of approach and probability of a nearby star in the globular cluster, of mass M, to induce the observed jerk in the system. }
    \label{Mass_bValues}
\end{figure}

\subsection{Developing a three-body model} \label{subsec: three-body model}

The precise measurements of $\dot{f}$, $\ddot{f}$, and $\dddot{f}$ for \psr\ allow for a characterisation of the external dynamical influence on the system. An attempt to fit for $f^{(4)}$ and $f^{(5)}$ in our timing solution revealed that they are not yet precisely measured and have large uncertainties, and this has prevented a complete model for this three-body problem. However, for our analysis, we have used the preliminary measurements to obtain further constraints on the solutions. An added uncertainty comes from the observed line-of-sight acceleration of the system, which is well within the maximum possible acceleration due to the globular cluster (as discussed above), and thus the exact contribution of the nearby star to $\dot{f}$ is not determinable. 

We developed a model to use the first three time derivatives of the pulse frequency for \psr\ and the limits on the fourth and fifth derivatives to derive the mass and orbital parameters of the external mass. With the caveats mentioned above, the dynamically-induced frequency derivatives solely due to the third body can be given by 
\begin{equation}
     f^{(n)} = -f\,\frac{\vec{a}^{(n-1)} \cdot \hat{\vec{K}}_0}{c} \,, 
\end{equation}
and $\vec{a}$ can be explicitly expressed using the orbital orientation of the binary and the eccentricity of the external orbit. We follow the method outlined in \citealt{Joshi_Rasio_1997} (hereafter \citetalias{Joshi_Rasio_1997}) and \cite{Perera_2017}, and the equations used in our analysis are presented in Appendix~\ref{sec:orbital_model}. The inferred orbital period of the external mass is much larger than that of the inner binary, and this allows us to make a good approximation of the system as the combination of two non-interacting Keplerian orbits \citep{TACL_1999}. The eccentric binary, currently the system \psr, is treated as a single object of mass $\mtot = \Mp + \Mc$, in an orbit with period $P_{\mathrm{b},M}$, eccentricity $e_{\mathrm{b},M}$, and semi-major axis $x_{M}$. The true anomaly, longitude of periastron, and orbital phase of the centre of mass of this binary with respect to the entire triple configuration are $\lambda_{M}$, $\omega_{M}$, and $\psi_{M}$. The outer companion has a mass $m_3$, and the orbital elements are $P_\mathrm{b,3}$ (for the case of an elliptical orbit), $e_3$, and $x_3$. The corresponding angles for the external mass are $\lambda_3$ and $\omega_3$, where $\lambda_3 = \lambda_{M}$ and $\omega_3 = \omega_{M} + 180\deg$. This external mass is moving at a velocity $v_3$ at a radial distance $r_3$ with respect to the centre of mass of the inner binary. 

We left the eccentricity of the external orbit ($e_3$) as a free parameter, and selected a value for $\lambda_{M}$ and $\psi_{M}$ for each trial (under the assumption of uniform prior probability distributions). Using the measurements of $\ddot{f}$ and $\dddot{f}$ and equations~\ref{eqn:f2_f3} and ~\ref{eqn:f4_f5}, we did an initial filtering of solutions to identify those where the contribution to the $\dot{f}$ from the external mass could be up to 5 times the observed value (to account for the acceleration of NGC~1851). We then calculated the estimated values for $f^{(4)}$ and $f^{(5)}$, and rejected further trials by comparing them with the measured limits. For our calculations, we have used $3\sigma$ upper limits on the  $f^{(4)}$ and $f^{(5)}$ values obtained from our timing analysis. This has allowed us to exclude the solutions for bound orbits with very short orbital periods and unbound orbits with very close distances of approach. For all the allowed solutions, we calculate $m_3$ and $a_3$ and derive the relevant parameters for the external mass.

\begin{figure}[!h]
    \centering
    \includegraphics[width=\hsize]{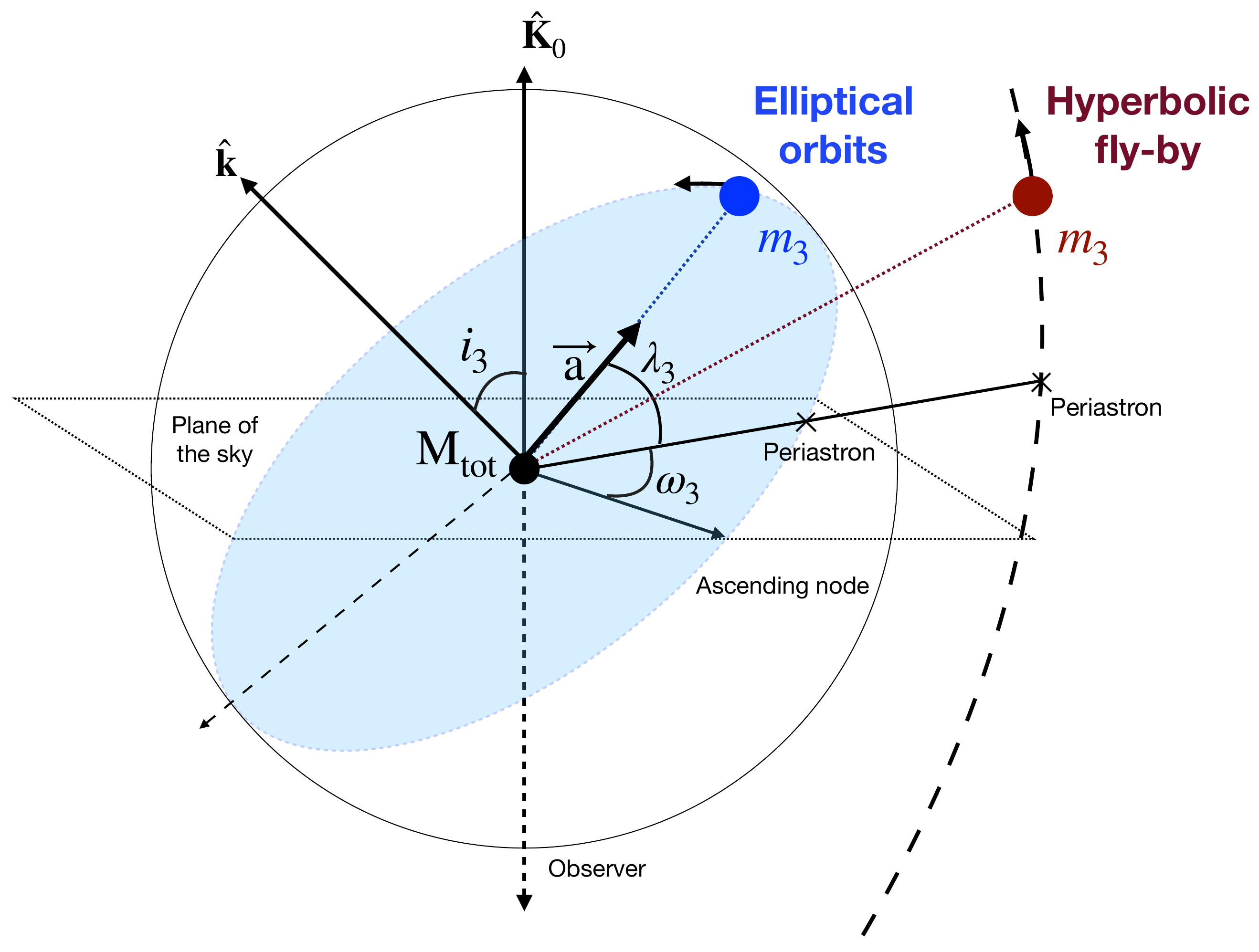}
    \caption{Definition of the orbital elements for a three-body model with two Keplerian orbits. The `inner binary' is represented by $\mtot$ and the external mass by $m_{3}$, and both the scenarios of an elliptical orbit and a hyperbolic fly-by encounter are shown.}
    \label{1851A_diagram}
\end{figure}

\subsection{Case I: A second mass in a hierarchical triple system}

We first consider the scenario where the system is in a hierarchical triple: a binary with the pulsar and its massive WD companion very close to each other, and the external third body as a second companion in a much longer orbit compared to the separation of the inner binary. 

Initially, we assume that the external orbit is circular ($e_3 = 0$), and attempt to derive a complete solution for the second companion. The measured first and third frequency derivatives of the pulsar can be used in equations (24) and (25) from \citetalias{Joshi_Rasio_1997} to directly derive $\lambda_M$ and $\dot{\lambda}_M$. However, the same signs of the derivatives produce an imaginary $\dot{\lambda}_{M}$, and straightaway eliminates the possibility of a circular orbit for the second companion. 

\begin{figure}[hbt!]
    \includegraphics[width=\columnwidth]{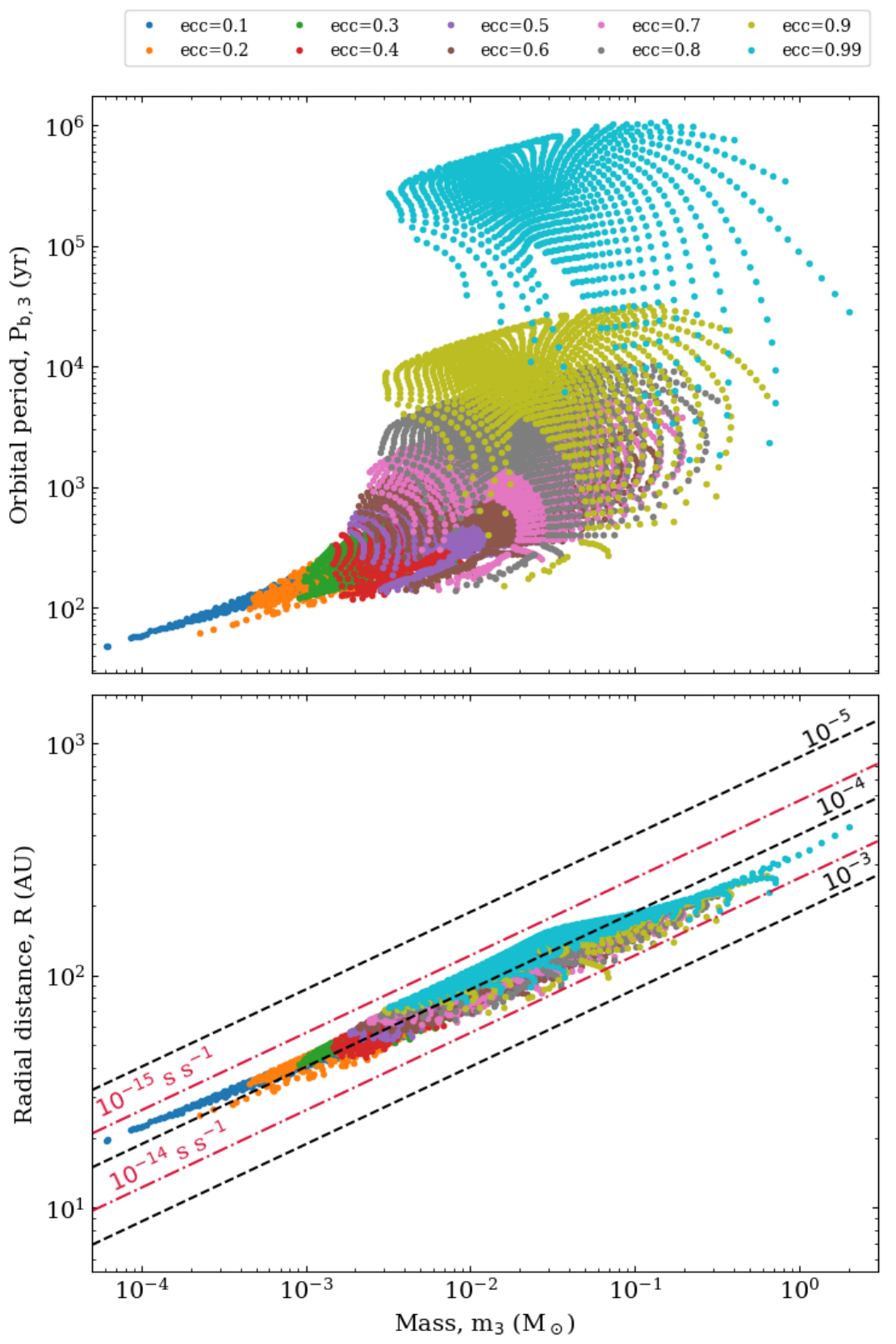}
    \caption{Family of solutions for the mass $m_3$,  orbital period $P_\mathrm{b,3}$, and relative distance $R$ for a second companion of \psr\, using the limits on all the frequency derivatives mentioned in the text. The different eccentricities are shown in different colours. The constant lines of $\dot\omega_{\mathrm{tidal}}/\dot\omega_{\mathrm{obs}}$ are shown with dotted black lines in the bottom plot, while the dash-dotted crimson lines indicate constant $\dot{x}_{\mathrm{tidal}}$.}
    \label{1851A_elliptical}
\end{figure}

We next applied our method described in the previous section to elliptical orbits for the second companion, and looked at a discrete range of eccentricities between 0.1 and 0.99, with the allowed space $0\,\leq\lambda_{M},\, \omega_{M},\,\psi_{M} < 360 \ \deg$. An initial analysis performed with only the values of $\dot{f}$, $\ddot{f}$ and $\dddot{f}$ suggested that the second companion could have a mass as low as $2.08\,\times\,10^{-10}\,\msun$ and an orbital period as small as $98$ days. To ensure consistency with the timing baseline of the system and its long-term stability, it is important that the second companion has a significantly larger orbit and does not come too close to the inner binary. This motivated us to further use the limits on $f^{(4)}$ and $f^{(5)}$ derived from the timing analysis, and we obtained a three-parameter family of solutions as illustrated in Fig.~\ref{1851A_elliptical}. 

For an elliptical outer orbit, we conclude that a second companion to \psr\ could have a mass $m_3$ between $6.6\,\times\,10^{-5}$ to 2.3 $\msun$, with an orbital period larger than 50 years and going up to several Myrs. Given the poor constraints on the current measurements of the higher frequency derivatives, the exact mass of the companion and the parameters of the outer orbit are undetermined. The nature of the objects that satisfy the allowed range of masses is pretty variable, ranging from sub-stellar objects like brown dwarfs and planets to stellar mass objects, neutron stars and  black holes.

\subsection{Case II: A hyperbolic encounter with a passing star}

We also considered the scenario where a hyperbolic encounter with a close-passing star is the cause for the remarkably large $\dot{a}_\mathrm{l}$ observed for \psr. A hyperbolic, or fly-by encounter, is a scenario when a nearby star passes by the target object in an unbound, hyperbolic trajectory. Such a close fly-by encounter involving an external star and a binary pulsar, capable of inducing such a large jerk in the latter, has not been previously observed. However, it has been suggested that clusters with high stellar density would be favorable hosts \citep{Mukherjee_2021}. This proposition for \psr\ is motivated by the exceptionally high central density of NGC~1851. The cluster also exhibits an intermediate rate of stellar interactions per binary, as evident by \psr\ itself being the product of a secondary exchange encounter. This parameter exponentially increases the likelihood of stellar encounters and close interactions. This makes the dynamical influence from a nearby star a likely possibility, given that the pulsar is located at a separation of 1\farcs58 from the core.

\begin{figure}[hbt!]
\centering
    \includegraphics[width=\columnwidth]{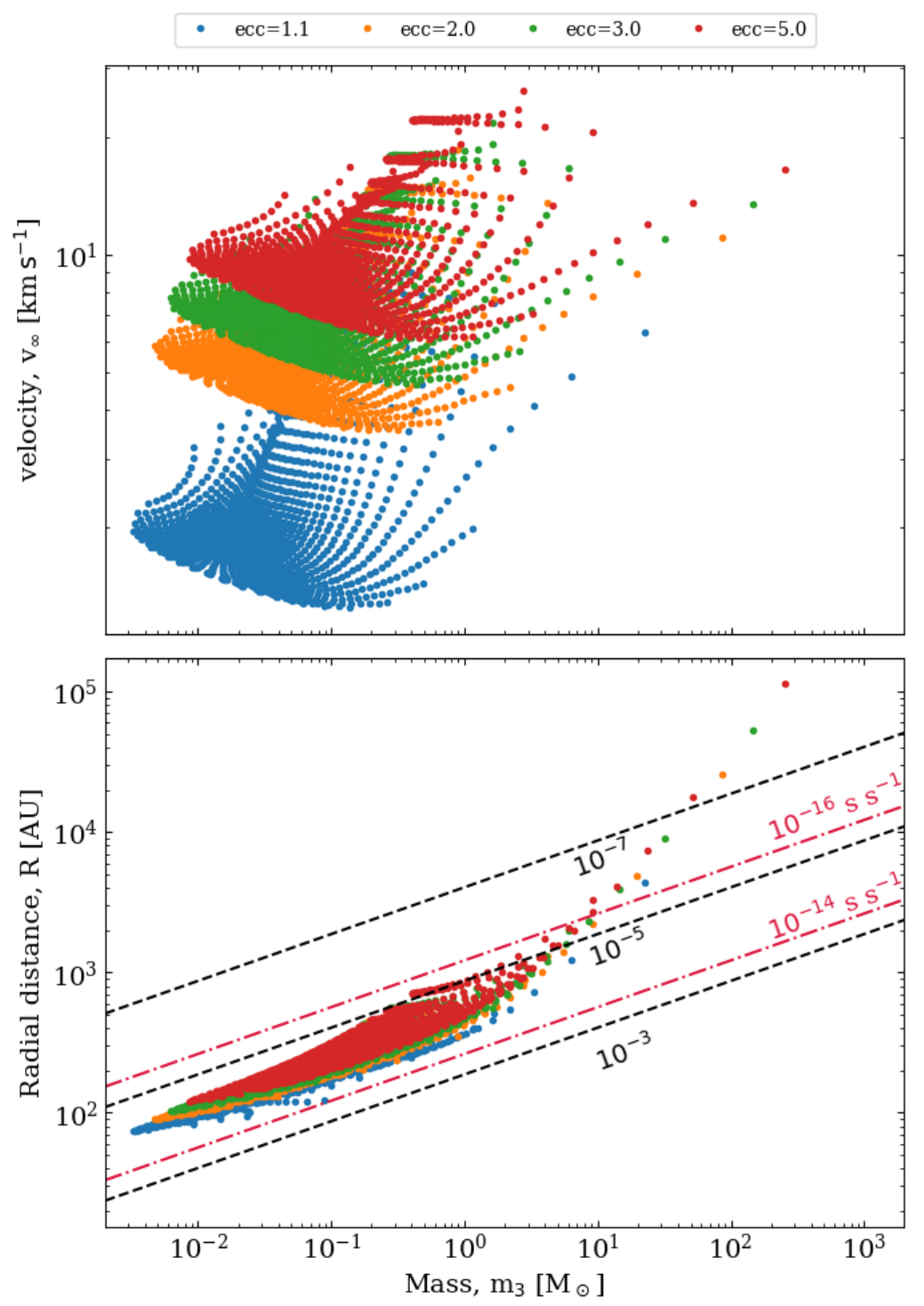}
    \caption{Allowed solutions showing the mass $m_3$, asymptotic velocity $v_\infty$, and relative distance $R$ for a nearby star in a fly-by encounter, in the core of the globular cluster. Constant $\dot\omega_{\mathrm{tidal}}/\dot\omega_{\mathrm{obs}}$ and constant $\dot{x}_{\mathrm{tidal}}$ are shown with black dotted and crimson dash-dotted lines respectively in the bottom plot.}
    \label{1851A_flyby}
\end{figure}

As detailed earlier, we varied both the $\lambda_{M}$ and $\psi_{M}$ values and analysed orbits with eccentricities ($e_3$) 1.1, 2.0, 3.0 and 5.0. Unlike a bound orbit, the true anomaly for a hyperbolic orbit varies between $-\pi + \cos^{-1}{\left(1/{e_3}\right)}$ and $\pi - \cos^{-1}{\left({1}/{e_3}\right)}$. For all eccentricities explored in this work, we find that the $\lambda_3$ is always less than 0, suggesting that the external mass is approaching the binary in the core of the globular cluster. We have implemented an additional criterion on the asymptotic velocity of the external mass ($v_\infty$), constraining it to be less than the central escape velocity of the cluster ($v_\mathrm{esc}$). With the increase in the eccentricity of the hyperbolic orbit, only the solutions with high velocity and larger distance are recovered. The allowed family of solutions are shown in Fig.~\ref{1851A_flyby}.

\subsection{Secular perturbations to the binary orbit}

It is very important to discuss the dynamical interactions for such a three-body system. We looked at the secular perturbations induced on the inner orbit due to the external mass, under the approximation that the latter has a fixed position in space with respect to the inner binary. These perturbative effects are mostly evident in the orbital eccentricity ($e$), the longitude of periastron ($\omega$), and the orbital period ($P_\mathrm{b}$). 

The changes in $P_\mathrm{b}$, due to this external mass, have been discussed in detail in the previous sections, and the changes in $e$ are not yet measurable. Here we make predictions for the change in the longitude of periastron of the inner binary due to the external mass ($\dot\omega_\mathrm{tidal}$) by calculating the orbital perturbation rates. For all possible orbital orientations, we estimate the maximum tidal contribution using the mass and radial separation of the external mass and the orbital parameters of the inner binary (equation~\ref{eq:maximum_omdot_tidal}). All detailed equations and calculations are presented in Appendix~\ref{sec:tidal_perturbations}.

In the bottom panels of Figs.~\ref{1851A_elliptical} and \ref{1851A_flyby}, we show the tidal contribution to the observed $\dot\omega$ value for bound and unbound orbits. For all eccentricities ranging from 0.1 to 5.0, $10^{-6} \leq \dot\omega_\mathrm{tidal}/\dot\omega_\mathrm{obs} \leq 1.3\times\,10^{-3}$, suggesting that the effect of the external body on the $\dot\omega_\mathrm{obs}$ could be an order of magnitude larger than the measurement uncertainty of $\dot{\omega}$, but this changes the total mass measurement by about $10^{-3}$. We can thus conclude that the $\dot\omega_\mathrm{obs}$ is mostly relativistic, and can be used for inferring the total mass of the system.

In these panels, we also see that the maximum $\dot{x}_\mathrm{tidal}$ is of the order of $1.4 \times 10^{-14} \, \rm s \, s^{-1}$, and normally much less than this value. Since, as mentioned above, this value
is correlated with the value of $\gamma$, it can also affect the mass measurements for the individual components of the binary, as discussed in the following section.

\subsection{Optical constraints on the nature of the companions of \psr}

Given the old age of NGC~1851 \citep[11.0 Gyr,][]{VandenBerg}, any stars still in the main sequence (MS) must have a mass of $\sim 0.8 \, \rm M_{\odot}$ or smaller: more massive stars will have long left the MS. The possible exception are blue straggler stars (BSS), which might have masses just up to twice that, $\sim 1.6 \, \rm M_{\odot}$.
From our calculations in the previous section, we conclude that any MS star must have a radial distance (R) from \psr\ smaller than about $800 \, \rm au$ (bottom panel in Fig.~\ref{1851A_flyby}). For a BSS with a higher mass, this distance does not change much. Given the parallax of this GC, 0.0857 mas, the region within R will have an angular radius of $\sim 0\farcs069$.

 \begin{figure*}
\centering
    \includegraphics[width=0.8\textwidth]{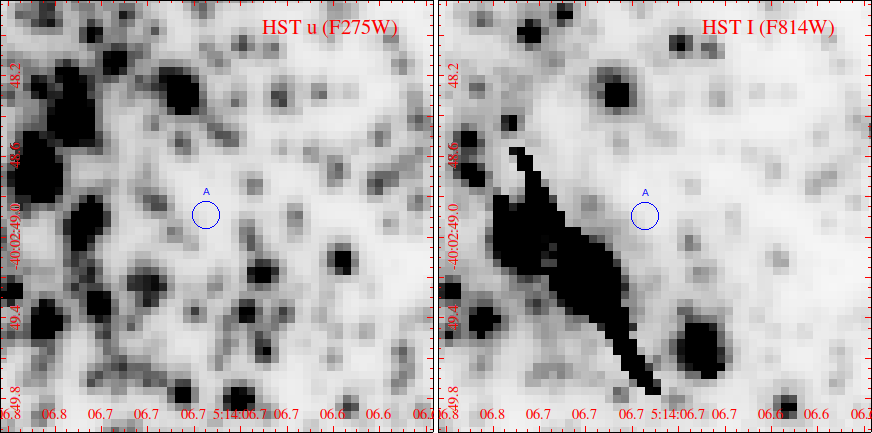}
    \caption{Location of \psr, with the blue circles corresponding to radius of 0\farcs069 around the position of the pulsar as estimated for the time the left frame was taken (MJD = 55558). Given the small proper motion, correcting for it does not induce any noticeable changes in the position of the pulsar. Images from the Wide Field Camera 3, taken with the F275W filter (left) and F814W filter (right).}
    \label{1851A_optical}
\end{figure*}

We have inspected archival Hubble Space Telescope (HST) images of NGC~1851; these are listed and described in detail by \cite{Barr_2024}. To locate the pulsar, we used the same astrometric solution that was used by the latter authors when searching the optical counterpart of NGC~1851E. The vicinity of the pulsar can be seen in Fig.~\ref{1851A_optical}, which includes part of an image taken with HST's Wide Field Camera 3, the blue circle representing the region of interest, within 0\farcs069 of \psr.  Given the pulsar's proximity to the cluster centre, where the crowdedness is severe, the faintest MS stars that can be detected in these images have masses of $\sim0.7 \, \msun$. 

As we can see, no sources are detected within the region of interest, either in the near-UV image on the left (taken with the F275W filter) or the near-IR image on the right (taken with the F814W survey). This excludes the possibility of the third star being a BSS, a giant or a MS star with a mass above $\sim0.7 \, \msun$.

Apart from obtaining constraints on the third star, these images also allow us to obtain definite conclusions regarding the massive binary companion of \psr. If it were an extended star, we would not be able to rely on the PK parameters to measure its mass. However, from the mass function we derived a lower limit for the companion mass of $0.84 \, \msun$  (Sect.~\ref{sec:keplerian}). Any BSS, MS or giant star with such a mass or larger should be clearly detectable at the position of the pulsar. This non-detection confirms that the companion of \psr\, is a compact object, either a WD or a NS. Such a conclusion was already expected from the lack of eclipses and DM variations in the timing data \citepalias{Ridolfi_19}.  

\section{Self-consistent analysis of masses and orbital orientation}\label{sec:chi2map}

The compact nature of the companion of \psr\, implies that the measured PK parameters quantify relativistic effects in the orbit, with small perturbations caused by the tides induced on the binary by the third object. This means that, within the uncertainty limits imposed by those perturbations, the GR equations of these PK parameters can be used to infer the masses of the components of the \psr\, system. We now proceed do do this.

\subsection{Bayesian map}

As discussed in detail by \citetalias{Ridolfi_19}, for wide binary systems, the effect of the Einstein delay ($\gamma$) is indistinguishable from a secular change of projected semi-major axis ($\dot{x}$). For this reason, we cannot separate both effects, but only measure their sum. Therefore, when estimating the masses from $\dot{\omega}$ and $\gamma$, we must take into account possible $\dot{x}$ contributions to our measurement of $\gamma$, which we can only estimate theoretically.

\citetalias{Ridolfi_19} did these calculations and came to the conclusion that by far the largest contribution to $\dot{x}$ is caused by the change of viewing angle that originates from the proper motion of the system \citep{Arzoumanian_1996,Kopeikin_1996}. This depends on the proper motion (with total magnitude $\mu$ and position angle $\Theta_{\mu}$), the unknown position angle of the line of nodes of the binary ($\Omega$) and the orbital inclination. Re-writing the formula given by \cite{Kopeikin_1996}, we obtain:
\begin{equation}
\dot{x}_{\mu} = x \mu \cot i \sin \left( \Theta_\mu - \Omega \right).
\end{equation}
Using our new estimate of the proper motion (2.76 mas yr$^{-1}$), and with $i \sim 60 \deg$, we obtain $| \dot{x}_{\mu}| \leq 0.9\, \times \, 10^{-14} \rm \, s \, s^{-1}$. This is smaller than estimated by \citetalias{Ridolfi_19}, both because of the smaller proper motion we measured, but also because of the more edge-on inclination. Nevertheless, the uncertainty in the value of $\gamma$ is also smaller, so that the effect of the proper motion on the measurement of $\gamma$ and in the mass estimates is still relevant.

However, the recognition that \psr\, has a nearby companion, and the current limits on its distance and mass imply $\dot{x}_\mathrm{tidal} = 1.4 \times 10^{-14} \, \rm s \, s^{-1}$, which is about 1.5 times larger than $\dot{x}_\mu$. However, this is an absolute maximum value: for most possible configurations (including the one that yields this maximum value), $\dot{x}_\mathrm{tidal}$ values are significantly smaller. So the assertion that $\dot{x}_{\mu}$ is the largest contribution to $\dot{x}$
is very likely correct.

These contributions must be compared to the apparent $\dot{x}$ that arises from $\gamma$ and its uncertainty, which according to
\citetalias{Ridolfi_19} is given by:
\begin{equation}
\dot{x}_\gamma = - \frac{\gamma \dot{\omega}}{\sqrt{1 - e^2}} \sin \omega,
\end{equation}
which, for the values in Table~\ref{table:timing_params}, results in $\dot{x}_\gamma = -30.0 \pm 0.7 \times 10^{-14}  \rm \, s \, s^{-1}$, which is more than 20 times larger than the maximum possible $\dot{x}_\mathrm{tidal}$.
We thus conclude that our measurement of $\gamma$ has at least that significance. Furthermore, the uncertainty on the measurement of
$\gamma$ is now smaller than the maximum value of $\dot{x}_{\mu}$; this means that the former no longer dominates the mass uncertainty and therefore the individual masses cannot be estimated much more precisely solely by improving $\gamma$.

\citetalias{Ridolfi_19} made a fully self-consistent estimate of the component masses and the orbital inclination of the system that takes the $\dot{x}_\mu$ fully into account. For this, they used the DDK orbital model, assumed the validity of GR, and made a map of the quality of fit ($\chi^{2}$) for every point in the orbital orientation space of a binary system with a known total mass $M_{\rm tot}$.

Since $\dot{x}_\mathrm{tidal}$ cannot yet be estimated properly, but is likely to be smaller than $\dot{x}_\mu$, the methodology proposed by \citetalias{Ridolfi_19} should still produce reliable uncertainty estimates. We followed that methodology by creating 2-D grids of the $\chi^{2}$ values under the consideration that randomly oriented orbits have a constant probability for $\cos i$. The model also intrinsically estimates all contributions from kinematic effects. 

The whole space ranges from $-1$ to 1 in $\cos i$ and 0 to 360 deg in $\Omega$: for each point in this grid, we hold the total mass ($M_{\mathrm{tot}}$) constant and introduce the corresponding values of $i$ and $\Omega$ in our modified DDK model using the KOM and KIN parameters. For each point in the given space, the masses of the pulsar and companion are calculated from the total mass and $i$ using equation~\ref{eq:mass_function}; the Einstein delay is calculated using equation~\ref{eq:gamma} respectively. These values are then added to the model using the M2 and GAMMA parameters respectively (the SINI parameter is defined automatically from KIN). The KOM, KIN, M2 and GAMMA parameters are kept fixed in the fit. Due to the contributions from various kinematic effects, we keep both $\dot{P}_{\mathrm{b}}$ and $\dot\omega$ as free parameters for generating the $\chi^{2}$ grids.

We then run TEMPO using the DDK model and allow for a fitting of all parameters not mentioned above. The post-fit $\chi^{2}$ value for each ($\cos{i}, \Omega$) combination is assigned to that specific grid point, and this value is then used to calculate the Bayesian 2D probability density function (PDF) at each point using \citep{Splaver_2002}:
\begin{equation}
    p(\Omega, \cos{i}) \propto \exp \left( {\frac{{\chi^2_\mathrm{min}}-{\chi^2}}{2}} \right)
\end{equation}
where ${\chi^2_\mathrm{min}}$ is the lowest $\chi^2$ of the whole grid. Following this, the 2D pdf is projected on the $\cos{i}$ and $\Omega$ axes and the corresponding 1D pdfs are calculated, which are shown in the upper left-hand and bottom right-hand panel of Fig. \ref{Full_chi2_map}. The 2D probability is shown in the central panel of the same figure.

\subsection{Results}

\begin{figure}[h!]
    % \centering
    % \includegraphics[scale=0.8]
    \includegraphics[width=\columnwidth]{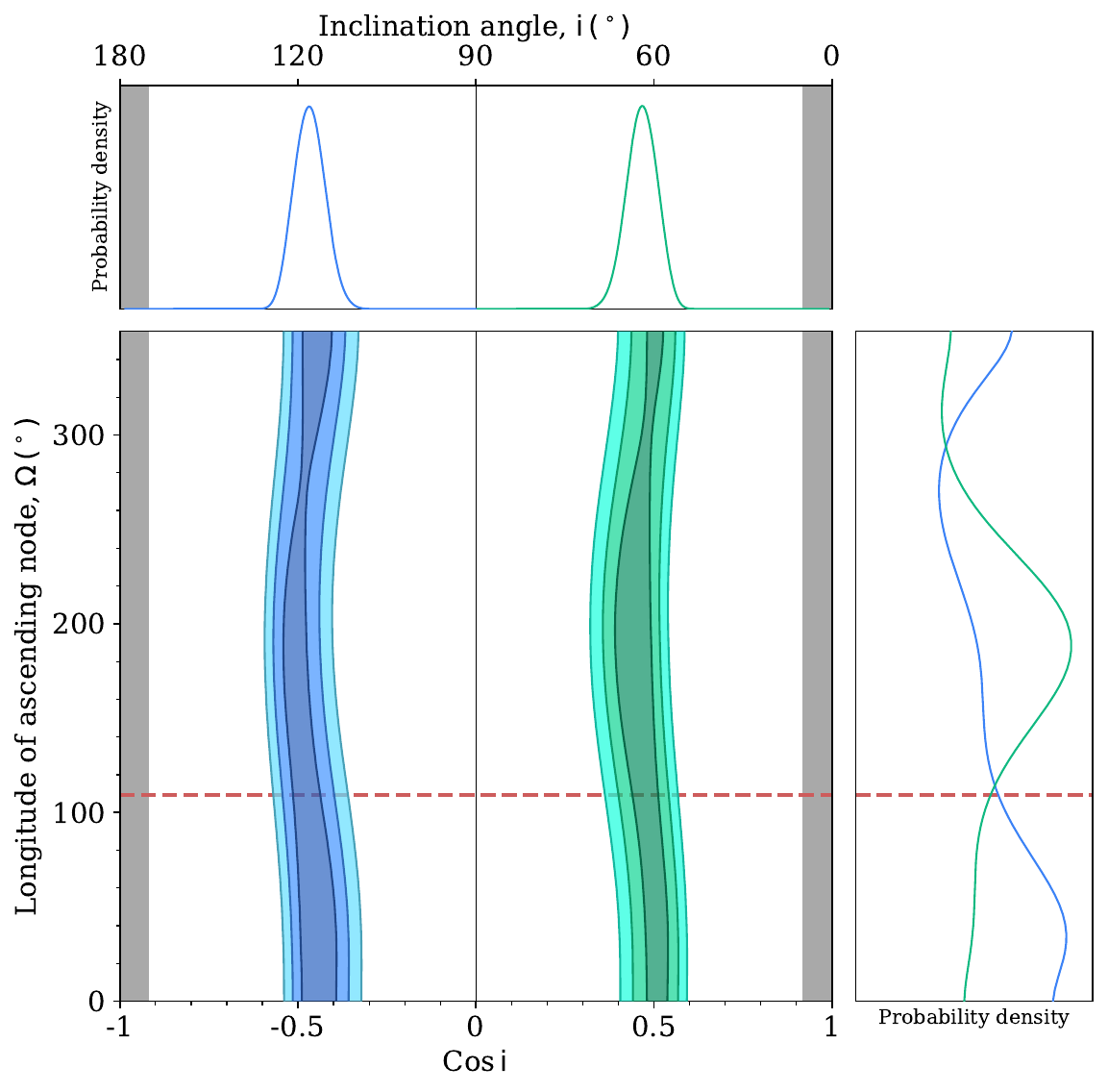}
    \caption{{Central panel}: 2-D probability density function (pdf) for the full $\Omega-\cos{i}$ space of binary pulsars. The blue and green contours show the probability density for inclination ${i}$ > $90\,\deg$ and ${i}$ < $90\,\deg$, where increasing probability density is shown with darker shades. The grey regions are excluded based on the requirement that the pulsar mass must be greater than 0, and the dotted red line indicates the position angle of the proper motion of the system. \textit{Top panel}: Probability density function for cos$\mathrm{i}$, normalised to the maximum. \textit{Right-hand panel}: Probability density function for $\Omega$.}
    \label{Full_chi2_map}
\end{figure}

It is evident from Fig.~\ref{Full_chi2_map} that the orbital inclination and $\Omega$ are correlated with each other. These variations of the best-fit values of $\cos i$ as a function of $\Omega$ are a clear indication that the effect of the proper-motion-induced $\dot{x}$ matters for this calculation. The identical probabilities for the two peaks of orbital inclination ($i = 62.6 \pm 3.0 \deg$, or $117.4 \pm 3.0 \deg$) shown in the top plot, is a result of a non-significant detection of the annual orbital parallax, the only effect that can be used to remove this degeneracy; this non-detection is expected given the very large distance to the pulsar. Because of the non-detection of the Shapiro delay, $\Omega$ is still not constrained. The derived masses for the pulsar and the companion, to the 68.3 percent confidence limit, are $\Mp = 1.39(3) \,\msun$ and $\Mc = 1.08(3)\,\msun$. This makes the pulsar mass 2.8-$\sigma$ larger and the companion mass 2.3-$\sigma$ smaller than what \citetalias{Ridolfi_19} measured, $M_{\mathrm{p}} = 1.25^{+0.05}_{-0.06}\,\msun$ and $M_{\mathrm{c}} = 1.22^{+0.06}_{-0.05}\,\msun$. The resulting lower companion mass now strongly indicates that the companion is a massive white dwarf. 

\section{Summary and Conclusions}
\label{sec:conclusions}
% \cM{Proper motion, position:} 
In this paper, we have presented the results of our updated timing analysis for \psr, using data from the GMRT and MeerKAT, over a period of $\sim$$18$ years. We have accounted for a 2-s offset in the GMRT dataset and extended the timing baseline of the system by more than 4 years since \citetalias{Ridolfi_19}. This has allowed for a very precise measurement of the position and proper motion of the system. The transverse velocity of the pulsar with respect to the cluster is $30\,(7)\,\mathrm{km}\,\mathrm{s}^{-1}$, and given the escape velocity of the cluster of $42.9\,\mathrm{km}\,\mathrm{s}^{-1}$, we can conclude that this is consistent with the pulsar's association to the cluster. 

% \cM{Spin and orbital frequency derivatives:} 
% \cM{Characteristic Age estimation:} 
We have obtained precise measurements of three spin frequency derivatives for the pulsar. Although the first derivative for pulsars in GCs is generally dominated by the acceleration from the cluster's mean field, the second derivative implies a line-of-sight jerk for \psr\, that is too high to be accounted by the latter. Additionally, the spin and orbital frequency derivatives show very large variations. In this paper we show that to estimate the orbital evolution correctly, we must take into account, higher order (> 1) orbital frequency derivatives, which are in this case estimated directly from the higher order spin frequency derivatives.

The inclusion of these higher derivatives and the aforementioned 2-s offset, in addition to the extension of the timing baseline, offered a precise measurement for $\dot{P}_{\mathrm{b,obs}} = 2.2\,(7)\,\times\,10^{-12}\,\mathrm{s}\,\mathrm{s}^{-1}$ at Epoch 2. This measurement, the estimate of the orbital period decay due to GW and the $\dot{P}_\mathrm{obs}$ measured at Epoch 2 allowed for a robust measurement of the pulsar's $\dot{P}_\mathrm{int}$, from which we derive a characteristic age of $4.8\,\mathrm{Gyr}$ and a surface magnetic field of $2.8\,\times\,10^8\,\mathrm{G}$.
The inclusion of the higher orbital frequency derivatives at Epoch 1 derived from the higher order spin period derivatives measured there leads to the derivation of 
different values of $\dot{P}_\mathrm{obs}$ and $\dot{P}_{\mathrm{b,obs}}$ at Epoch 1, but from them consistent values of
$\dot{P}_\mathrm{int}$ can be obtained. 

% \cM{External body:}
The large second spin frequency derivative, we argue, is due to the presence of a third body near the binary, and both elliptical and hyperbolic orbits are possible.
Although we cannot yet conclude the exact nature of this object, we have derived a family of solutions for the orbital parameters for different orbital eccentricities. This family of solutions indicates that this third object, if a main sequence star, must have a radial distance smaller than $800$ au from \psr. This, in combination with optical images of the cluster, exclude the possibility of the third object being a blue straggler star, or a main sequence star more massive than $0.7\, \msun$. If this system is in a hierarchical triple, it would allow an insight into the formation of such triples in globular clusters. On the other hand, a hyperbolic encounter will make this system the first of its kind and provide a test bed for studying the dynamical interactions in this cluster. Further long-term timing measurements will offer tighter constraints on the higher frequency derivatives, and potentially distinguish between ellipsoidal and hyperbolic orbits. 

The absence of any bright star at the position of \psr\, in HST images of NGC~1851 confirms that the binary companion to \psr\, is a compact object, either a NS or a massive WD. The possibility of an extended companion to \psr\, had been advanced earlier because of the long-timescale scintillation seen during different orbital phases of the pulsar, which could have been caused by gas emanating from the companion \citep{Freire_2007}. With the discovery of a handful of other pulsars in this cluster by \cite{Ridolfi_2022} and the observation of similar scintillation for all of them (including the isolated ones), we can conclude that the observed scintillation of \psr\ is likely not due to its companion.

In this work, we have obtained improved and robust measurements of the $\dot\omega$ and $\gamma$ parameters for the system. The contribution to the $\dot\omega$ from the third body is at least 3 orders of magnitude smaller than the observed value, and thus we conclude that the observed $\dot\omega$ is almost purely relativistic.
The maximum possible value of $\dot{x}_\mathrm{tidal}$ could change $\gamma$ by a factor of up to 1/20;
however this number is likely much smaller, so we conclude that the observed $\gamma$ is also mostly relativistic.
The masses derived from a self-consistent analysis of the timing, which take into account the contribution of the proper motion to $\dot{x}$ and $\gamma$, are $\mtot = 2.47346\, \msun$ (the latter has a relative uncertainty due to 
$\dot{\omega}_\mathrm{tidal}$ that could be up to $10^{-3}$), $\Mp = 1.39(3)\,\msun$ and $\Mc = 1.08(3)\,\msun$. The latter are 2.8 and 2.3-$\sigma$ different from the masses obtained by \citetalias{Ridolfi_19} respectively, $M_{\mathrm{p}} = 1.25^{+0.05}_{-0.06}\,\msun$ and $M_{\mathrm{c}} = 1.22^{+0.06}_{-0.05}\,\msun$, the reason for this is the decrease in the value of $\gamma$ that has happened with the inclusion of more data. Given the recent stabilization in the value of $\gamma$, we deem these mass values to be more reliable than those published by \citetalias{Ridolfi_19}.
With the new mass estimate for the companion derived in this work, and considering the currently lightest known NS mass of $1.17\,\msun$ \citep{Martinez_2015}, it becomes much more likely that the companion of \psr\ is a massive white dwarf.

%%%%%%%%%%%%%%%%%%%%%%%%%%%%%%%%%%%%%%%%%%%%%%%%%%%%%%%%%%%%%%%%%%%%%%%%%%%%%%%%

\begin{acknowledgements}
The MeerKAT telescope is operated by the South African Radio Astronomy Observatory, which is a facility of the National Research Foundation, an agency of the Department of Science and Innovation. SARAO acknowledges the ongoing advice and calibration of GPS systems by the National Metrology Institute of South Africa (NMISA) and the time– space reference systems department of the Paris Observatory. MeerTime data is housed on the OzSTAR supercomputer at Swinburne University of Technology. The OzSTAR program receives funding in part from the Astronomy National Collaborative Research Infrastructure Strategy (NCRIS) allocation provided by the Australian Government. MeerKAT observations used the PTUSE and APSUSE computing clusters for data acquisition, storage, and analysis. These clusters were funded and installed by the Max- Planck-Institut für Radioastronomie (MPIfR) and the Max- Planck-Gesellschaft. PTUSE was developed with support from the Australian SKA Office and Swinburne University of Technology. The authors also acknowledge Max-Planck-Institut f\"{u}r Radioastronomie funding to contribute to MeerTime infrastructure. We thank the staff of the GMRT that made the observations possible. GMRT is run by the National Centre for Astrophysics of the Tata Institute of Fundamental Research. The National Radio Astronomy Observatory is a facility of the National Science Foundation operated under cooperative agreement by Associated Universities, Inc. 
\\
AD, PCCF, TG, NW, DJC, MK, VVK, VB, and PVP acknowledge continuing valuable support from the Max-Planck Society. FA acknowledges that part of the research activities described in this paper were carried out with the contribution of the NextGenerationEU funds within the National Recovery and Resilience Plan (PNRR), Mission 4 – Education and Research, Component 2 – From Research to Business (M4C2), Investment Line 3.1 – Strengthening and creation of Research Infrastructures, Project IR0000034 – 'STILES -Strengthening the Italian Leadership in ELT and SKA'. MBa acknowledges support through ARC grant CE170100004. YG acknowledges support from the Department of Atomic Energy, Government of India, under project numbers 
12-R\&D-TFR-5.02-070 and RTI4002. AP acknowledges that this work was supported in part by the "Italian Ministry of Foreign Affairs and International Cooperation", grant number ZA23GR03, under the project "RADIOMAP- Science and technology pathways to MeerKAT+: the Italian and South African synergy". SMR is a CIFAR Fellow and is supported by the NSF Physics Frontiers Center award 1430284. LZ is supported by ACAMAR Postdoctoral Fellowship and the National Natural Science Foundation of China (Grant No. 12103069). Finally, we appreciate the insightful comments and suggestions provided by the anonymous referee on the manuscript. 
\end{acknowledgements}

%%%%%%%%%%%%%%%%%%%%%%%%%%%%%%%%%%%%%%%%%%%%%%%%%%%%%%%%%%%%%%%%%%%%%%%%%%%%%%%%

\bibliography{main.bib}{}
\bibliographystyle{aa}

%%%%%%%%%%%%%%%%%%%%%%%%%%%%%%%%%%%%%%%%%%%%%%%%%%%%%%%%%%%%%%%%%%%%%%%%%%%%%%%%
%%%%%%%%%%%%%%%%%%%%%%%%%%%%%%%%%%%%%%%%%%%%%%%%%%%%%%%%%%%%%%%%%%%%%%%%%%%%%%%%

\appendix

%%%%%%%%%%%%%%%%%%%%%%%%%%%%%%%%%%%%%%%%%%%%%%%%%%%%%%%%%%%%%%%%%%%%%%%%%%%%%%%%

\section{GMRT clock offsets}
\label{sec:appendix_GMRT_clock}

Two years after its release with 16 antennas, the wideband correlator (GWB) of the GMRT was upgraded to a 32$-$antenna version in 2017. Following this upgrade, the actual sampling time of the recorded data had a time difference with the accurate reference timestamps of observations \footnote{Reddy et al, \href{http://www.gmrt.ncra.tifr.res.in/subsys/digital/DigitalBackend/target_files/GWB/ITR/GWB_Timestamp_ITR_Release_1.0.pdf}{NCRA Internal Technical Report}, April 2022}. For the beam data acquired with the GMRT, this difference was constant. The exact values for the phased array (PA) and the coherent de-dispersion (CD) beams, for different bandwidths (BWs) and between corresponding dates, are noted in tables \ref{first} and \ref{second} below respectively. All differences were corrected for beam data acquired after 14 May 2020. We include these values here for easy reference.

\vspace{-0.05in}
\setcounter{totalnumber}{2}
\begin{table}[H]
  \renewcommand{\thetable}{\arabic{table}a}
   \caption{Offsets for PA beam data}\label{first}
     \vspace{-0.05in}
  \begin{tabular}{|c|c|c|}
   \hline
    Dates & 200/\:400 MHz BW & $\leq$ 100 MHz BW\\
    \hline
    14/08/2015 - 09/09/2016 & 0.17096239 & 0.84190767\\
    \hline                        
    09/09/2016 - 14/05/2020 & 1.34217728 & 2.68435456\\
    \hline
    \end{tabular}
%   \caption{First caption}\label{first}
\end{table}
\vspace{-0.25in}
\begin{table}[H]
\addtocounter{table}{-1}
  \renewcommand{\thetable}{\arabic{table}b}
  \caption{Offsets for CD beam data}\label{second}
  \vspace{-0.05in}
  \begin{tabular}{|c|c|c|}
   \hline
    Dates & 200/\:400 MHz BW & $\leq$ 100 MHz BW\\
     \hline                        
    04/07/2017 - 14/05/2020 & 2.01326592 & 4.02653184\\     
    \hline
    \end{tabular}
    %\tablefoot{BW:Bandwidth}
\end{table}
\vspace{-0.02in}

%%%%%%%%%%%%%%%%%%%%%%%%%%%%%%%%%%%%%%%%%%%%%%%%%%%%%%%%%%%%%%%%%%%%%%%%%%%%%%%%

\section{Stability of measured post-Keplerian parameters}
\label{sec:PK_parameters_stability}

\begin{figure}[!h]
    \centering
    \includegraphics[width=\hsize]{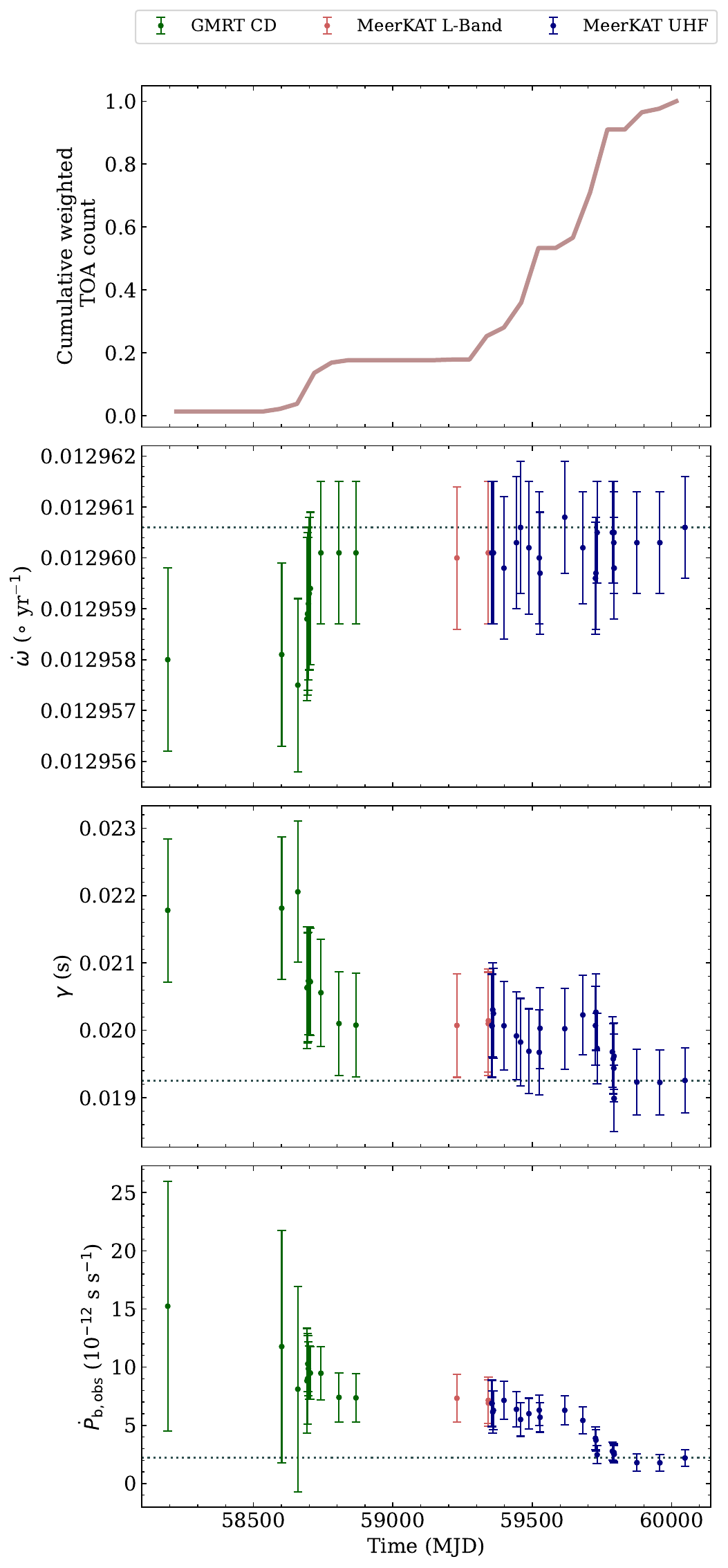}
    \caption{\textit{Top to bottom}: Cumulative weighted TOA count, observed rate of advance of periastron $\dot\omega_{\mathrm{obs}}$, Einstein delay $\gamma$, and observed orbital period derivative $\dot{P}_{\mathrm{b,obs}}$ as a function of time. The values and their residual uncertainties (given by the vertical errorbars) are as measured using the DDH timing model presented in Table~\ref{table:timing_params}.}
    \label{parameter_evolution}
\end{figure}

We have also done self-consistent checks for the stability of the observed $\dot\omega$, $\gamma$, and $\dot{P}_{\mathrm{b,obs}}$ with the extension of the dataset; this is shown in figure~\ref{parameter_evolution}. The measured values of the parameters are consistent throughout the dataset and the fluctuations are all consistent within the $1\,\sigma$ uncertainties. The stability of the measurements indicate that the constraints on the masses are reliable and there are no bad observations that might influence a biased measurement, suggesting that they will not change significantly in the future. 

An observing campaign for NGC~1851 during July-August 2023 \citep{Barr_2024} produced a dense dataset for \psr, consisting of 18 hours of data across six observations with the UHF receiver of MeerKAT. These high-cadence observations improved the sampling and orbital coverage of the TOAs, resulting in precise and stable measurements of the parameters. As evident from the bottom panel of Fig.~\ref{parameter_evolution}, a robust measurement for $\dot{P}_\mathrm{b,obs}$ was obtained at the end of the observing campaign, which remains stable with the addition of more data. 

%%%%%%%%%%%%%%%%%%%%%%%%%%%%%%%%%%%%%%%%%%%%%%%%%%%%%%%%%%%%%%%%%%%%%%%%%%%%%%%%

\section{Three-body orbital model}
\label{sec:orbital_model}

We used equations (11) - (13) given by \citetalias{Joshi_Rasio_1997}, and an extension of the same by \citealt{Perera_2017} (Appendix A) to derive the parameters of the external mass using the explicit expressions for the frequency derivatives of the pulsar. Unlike them, we used the $\ddot{f}$ as our primary measurement, and propagated the dependence of the other derivatives on the same. All the equations have a dependence on the sum of $\lambda$ and $\omega$, the true anomaly and the longitude of periastron respectively. We define the parameter $\psi$ as $\psi = \lambda + \omega$, and a schematic of the geometry is shown in Fig.~\ref{1851A_diagram}.

The orbital parameters of the centre of mass of the binary are given with a subscript `$M$' and those of the external mass with a `3', with respect to the centre of mass of the entire three-body configuration. In addition to the three frequency derivatives precisely measured for \psr, we have also obtained constraints on the fourth and fifth derivatives. These derivatives, as purely induced by the external dynamics, after the removal of kinematic and intrinsic contributions, can be given as 
\begin{align}
  &\dot{f}   = -f\,\frac{\vec{a} \cdot \hat{\vec{K}}_0}{c} \,, \label{f1} \\  
  &\ddot{f}  = -f\,\frac{\dot{\vec{a}}\cdot \hat{\vec{K}}_0}{c} \,, \label{f2} \\
  &\vdots \nonumber\\
  &{f}^{(5)} = -f\,\frac{\vec{a}^{(4)}\cdot \hat{\vec{K}}_0}{c} \,. \label{f5} 
\end{align}

The terms on the right can be explicitly expressed using the orbital parameters of the binary and the angular separations. Here we present these expressions, derived using the method outlined by \citetalias{Joshi_Rasio_1997}, and transformed to a function of $\psi$ and $\dot\psi$. These equations are valid for both elliptical and hyperbolic orbits, and can be solved to derive the unknown parameters for the third body. 

The altered  equations used in our code for obtaining the family of solutions presented earlier are

\begin{alignat}{1000}
    % &\dot{f} = -fK{A^2}\sin{\psi} \\
    {f}^{(2)}& = \frac{{B}{\dot\psi_{M}}{\dot{f}}}{{A^2}{\sin{\psi_{M}}}}\,,\quad & \dddot{f} & = \frac{{C}{{\dot\psi_{M}}^2}{\dot{f}}}{{A^2}{\sin{\psi_{M}}}} \,,\label{eqn:f2_f3} \\ 
    {f}^{(4)}& = \frac{{D}{{\dot\psi_{M}}^3}{\dot{f}}}{{A^2}{\sin{\psi_{M}}}}\,,\quad & {f}^{(5)}& = \frac{{E}{{\dot\psi_{M}}^4}{\dot{f}}}{{{A}^2}{\sin{\psi_{M}}}} \,, \label{eqn:f4_f5}
\end{alignat}
where a prime indicates a derivative with respect to $\psi_{M}$, and $A$, $B$, $C$, $D$, and $E$ \citep{Joshi_Rasio_1997, Perera_2017} are defined as 
\begin{align}
    A &= 1 + e_3\,\cos\lambda_M \,,\\
    B &= 2{A}{A'}\sin{\psi_{M}} + A^{2}\cos{\psi_{M}} \,, \\
    C &= B' + \frac{2{B}{A'}}{A}\,,\;
    D = C' + \frac{4{C}{A'}}{A}\,,\;
    E = D' + \frac{6{D}{A'}}{A}\,.
\end{align}
We derive $\dot\psi$ using the equations \ref{eqn:f2_f3}, and use it to derive $\dot{f}$, ${f}^{(4)}$, and ${f}^{(5)}$ from ${f}^{(2)}$, using the equations
\begin{alignat}{10}
    {\dot\psi_{M}} & = \left(\frac{\dddot{f}}{{f}^{(2)}}\right)\left(\frac{B}{C}\right) \,, \\
    {\dot{f}}      & = \frac{{{A^2}}}{{B}{\dot\psi_{M}}}{{f}^{(2)}}{\sin{\psi_{M}}}\,,\;\; 
        & {f^{(4)}} & = \frac{{D}{B}}{{\dot\psi_{M}}^2}\,,\;\; 
        & {f^{(5)}} & = \frac{{E}{B}}{{\dot\psi_{M}}^3}\,.
\end{alignat}
%%%
The calculated values of the frequency derivatives are compared with the observed limits, and the allowed solutions for $\lambda$, $\psi_{M}$, and $\dot\psi_{M}$ are used to determine the parameters of the companion using equations (14)-(18) from \citetalias{Joshi_Rasio_1997}.

For an external third body in an elliptical orbit, the mass ($m_3$) and semi-major axis ($a_3$) are given by 
\begin{align}
    &{m_3}\sin{i_3} \approx -\frac{\,\dot{f}{c}}{f\sin\psi_{M}}\left[\frac{{\mtot^2}{A^2}}{G{\dot\psi_{M}^{4}}}\right]^{1/3} \label{m3}\\
    &{a_3} = \left(\frac{\mtot}{m_3}\right)a_{M}  =
    \left(\frac{\mtot}{m_3}\right) \frac{h}{1-e_3^{2}}, \label{a_3}\\
    &h = -\frac{\dot{f}{c}{{\rm A}^2}}{{f}\,{\sin{i_3}}\,{\sin\psi_{M}}\,{{\dot\psi_{M}}^2}} \label{h}, 
\end{align}
where $h$ is the semi-latus rectum of the orbit of the binary's centre of mass, and the semimajor axis of the relative orbit between $\mtot$ and $m_3$ is $a_R = a_M + a_3$. The relative separation between the two ($R$), the orbital frequency and orbital period of the orbit of the external mass ($n_{\rm b,M}$, $P_{\rm b,M}$), and its (relative) velocity in this orbit ($v$) are given by \citep{Roy_OrbitalMotion}
\begin{alignat}{100}
    &{R} = \frac{{a_{R}}{(1-e_3^{2})}}{1+e_3\cos{\lambda_M}} \,, \\
    &{n_{\mathrm{b},M}} = \left[\frac{G{(\mtot +m_3)}}{a_R^{3}}\right]^{1/2}, & P_{\rm b,M} = \frac{2 \pi}{n_{\mathrm{b},M}} \,,\\
    &{v} = \left[G(\mtot +m_3)\left({\frac{2}{R} - \frac{1}{a_{R}}}\right)\right]^{1/2}\,.
\end{alignat}

For the external body in a hyperbolic orbit, the mass $m_3$ can be derived using equation~\ref{m3}; however, the semi-major axis $a_3$, velocity $v$ and relative separation $R$ of this body are given as \citep{Roy_OrbitalMotion}
\begin{align}
    &{a_3} = \left(\frac{\mtot}{m_3}\right)a_{M}  =
    \left(\frac{\mtot}{m_3}\right) \frac{h}{e_3^{2}-1}, \label{a_3_hyp}\\
    &v = \left[G(\mtot +m_3)\left({\frac{2}{R} + \frac{1}{a_{R}}}\right)\right]^{1/2}\\
    &{R} = \frac{{a_{R}}{(e_3^{2} -1)}}{1+e_3\cos{\lambda_M}} \,.
\end{align} 
The asymptotic velocity of this body at infinity can be further derived as 
\begin{equation}
    v_\infty = \left[\frac{G(\mtot +m_3)}{a_{R}}\right]^{1/2}\,.
\end{equation}
Since the system is present in the core of the GC, we additionally ensure, for each trial, that the asymptotic velocity of the the external mass, $v_\infty$, is less than twice the central escape velocity of the cluster. As discussed earlier, we also put limits on $\psi_{M}$ according to the eccentricity $e_3$ of the external mass' trajectory. 

%%%%%%%%%%%%%%%%%%%%%%%%%%%%%%%%%%%%%%%%%%%%%%%%%%%%%%%%%%%%%%%%%%%%%%%%%%%%%%%%

\section{Perturbations of the binary orbit by the third body}
\label{sec:tidal_perturbations}

The presence of a distant third body leads to tidal perturbations of the pulsar binary. It is important to quantify these perturbations and estimate their relevance for the conclusions of the main part of the paper, most importantly the total mass of the system derived under the assumption that the observed $\dot\omega$ can be interpreted as the relativistic periastron advance of a binary system. In this appendix we summarise the calculations needed to estimate the secular changes in the orbital parameters due to the perturbations by a distant  third mass $m_3$. The orbital parameters for the orbit of this third body are taken from Sect.~\ref{sec:orbital_model}.

Using the same convention as in Sect.~\ref{sec:results}, we describe the geometry of the binary orbit (relative motion $\vec{r} = \vec{r}_\mathrm{p} - \vec{r}_\mathrm{c}$) by the following Keplerian parameters:
\begin{itemize}
\item $a$: semimajor axis of the relative binary motion 
\item $e$: orbital eccentricity
\item $i$: orbital inclination with respect to the plane of the sky, i.e.\ the angle between the orbital angular momentum (direction $\hat{\vec{k}}$) and the line of sight from the observer to the pulsar binary (direction $\hat{\vec{K}}_0$)
\item $\Omega$: longitude of the ascending node (direction $\hat{\vec{i}}$), measured from a fixed direction in the plane of the sky, e.g.\ North (clockwise as seen by the observer).
\item $\omega$: angular distance of the periastron from the ascending node. In pulsar astronomy this angle is called the {\it longitude of periastron}.\footnote{Note, in celestial mechanics, the name {\it longitude of periastron / pericentre / periapsis} is often used for $\varpi := \Omega + \omega$. The angle $\omega$ is then called the {\it argument of periastron / pericentre / periapsis}. See for instance \cite{Danby_book}.}
\end{itemize}
In the following, we use a coordinate system $XYZ$ where the $X$-$Y$ plane is the plane of the sky and $Z$ is in the direction away from the observer ($\hat{\vec{K}}_0$). Without loss of generality we can assume the $X$-axis to be aligned with the ascending node of the binary orbit at the epoch $t_0$ (practically identish to $T_0$). Consequently, the unit vectors related to the orbital plane as defined by \cite{Damour_Taylor_1992} are given by
\begin{equation}
  \hat{\vec{i}} = (1,0,0) \,,\;
  \hat{\vec{j}} = (0, \cos i,\sin i) \,,\;
  \hat{\vec{k}} = (0,-\sin i,\cos i) \,.
\end{equation}

As a first approximation, which is sufficient for the purposes of this paper, we can assume in the following that the perturbing third body has a fixed direction in space. The direction from the centre of mass of the binary to the third body at a distance $R$ is given by the unit vector $\hat{\vec{N}}$. Using standard spherical coordinates $(\varphi,\vartheta)$ with respect to $XYZ$ one has
\begin{equation}
  \hat{\vec{N}} = (\sin\vartheta_3\cos\varphi_3, \, \sin\vartheta_3\sin\varphi_3, \, \cos\vartheta_3) \,.
\end{equation}
From Sect.~\ref{sec:orbital_model} one has
\begin{equation}
  R = \frac{a_R(1 - e_3^2)}{1 + e_3 \cos \lambda_M} \,,
\end{equation}
where $a_R$ is the semimajor axis of the relative orbit between the third body and the centre of mass of the binary system. Furthermore, one has (cf.\ Sect.~\ref{sec:orbital_model})
\begin{equation}
  \vartheta_3 = \arccos(N_Z) 
              = \arccos(\sin\psi_3 \sin i_3) \,.
\end{equation}
where $\psi_3 = \omega_3 + \lambda_M$. Note that the azimuth-angle $\varphi_3$ of the third body is unconstrained by the frequency derivatives (cf. Sect.~\ref{sec:orbital_model}) and therefore needs to be treated as a free parameter. 

In the following we need the direction cosines
\begin{align}
& \xi_{\hat{\vec{i}}} := \hat{\vec{N}} \cdot \hat{\vec{i}} = \sin\vartheta_3\cos\varphi_3 \,,\\
& \xi_{\hat{\vec{j}}} := \hat{\vec{N}} \cdot \hat{\vec{j}} = \cos\vartheta_3\sin i +  \sin\vartheta_3\sin\varphi_3\cos i \,, \\
& \xi_{\hat{\vec{k}}} := \hat{\vec{N}} \cdot \hat{\vec{k}} = \cos\vartheta_3\cos i -  \sin\vartheta_3\sin\varphi_3\sin i \,.
\end{align}

The perturbation of a binary orbit by a distant third body can be calculated using the formalism of osculating elements (see for instance \ Sect.~3.4.1 in the book by \cite{Poisson_Will_book}). Integrating over one full binary orbit one obtains the following shifts in the orbital parameters:
\begin{align}
&\Delta a = 0 \,, \label{eq:Da3} \\
&\Delta e = \frac{15\pi \eta}{2} \, e \sqrt{1-e^2} 
\nonumber\\ 
&\qquad \times \left[
    (\xi_{\hat{\vec{i}}}^2-\xi_{\hat{\vec{j}}}^2) \, \sin (2 \omega )
    -2\,\xi_{\hat{\vec{i}}} \,\xi_{\hat{\vec{j}}} \, \cos (2 \omega )
  \right] \,, \label{eq:De3} \\
&\Delta i =\! \frac{3\pi\eta}{2} \, \frac{\xi_{\hat{\vec{k}}}}{{\sqrt{1-e^2}}} \nonumber\\
&\qquad \times \, \left[ 
(2 + 3 e^2) \, \xi_{\hat{\vec{i}}}
+ 5 e^2 \xi_{\hat{\vec{i}}} \cos (2 \omega )
+ 5 e^2 \xi_{\hat{\vec{j}}} \sin (2 \omega) 
\right] \,, \label{eq:Di3}  \\
&\Delta\Omega = \frac{3\pi\eta}{2} \,\frac{\xi_{\hat{\vec{k}}}}{{\sqrt{1-e^2} \sin i}} 
\nonumber\\ 
&\qquad \times \, \left[ 
  (2 + 3 e^2) \, \xi_{\hat{\vec{j}}}
  + 5 e^2 \xi_{\hat{\vec{i}}} \sin (2 \omega )
  - 5 e^2 \xi_{\hat{\vec{j}}} \cos (2 \omega) 
  \right] \,, \label{eq:DOm3} \\
&\Delta\omega = \frac{3\pi\eta}{2} \sqrt{1 - e^2} 
\nonumber\\ 
&\qquad \times \left[
  1 - 3 \xi_{\hat{\vec{k}}}^2 
  + 5 (\xi_{\hat{\vec{i}}}^2-\xi_{\hat{\vec{j}}}^2) \, \cos (2 \omega ) 
  + 10 \,\xi_{\hat{\vec{i}}} \xi_{\hat{\vec{j}}} \sin(2\omega ) \right] 
\nonumber\\ 
&\qquad   - \cos i\,\Delta\Omega \,, \label{eq:Dom3}
\end{align}
where  

\begin{equation}
    \eta = \frac{m_3}{\mtot} \left(\frac{a}{R}\right)^3 \ll 1 \,.
\end{equation}

These equations agree with the ones obtained by \cite{Lidov_1962} (cf.\ his Eqs.~17).\footnote{Note, \cite{Lidov_1962} uses a different set of direction cosines. They are related to the ones used in this paper by  
$\xi_1 =  \xi_{\hat{\vec{i}}}\cos\omega + \xi_{\hat{\vec{j}}}\sin\omega$, 
$\xi_2 = -\xi_{\hat{\vec{i}}}\sin\omega + \xi_{\hat{\vec{j}}}\cos\omega$, and
$\xi_3 =  \xi_{\hat{\vec{k}}}$. } 
The corresponding (secular) rates of change of the orbital parameters are simply obtained by dividing the parameter shifts by the orbital period, for example $\dot{\omega} \equiv \Delta\omega/P_\mathrm{b}$.

As equation~(\ref{eq:DOm3}) suggests, the secular change in the longitude of periastron depends on the orientation of the third body and the inclination, eccentricity, and longitude of periastron of the inner binary. For the purpose of this work, it is sufficient to estimate the maximum expected tidal perturbation to the observed rate of advance of periastron, $\dot\omega_{\rm tidal,max}$, over all possible directions of the external third body. Using the estimated orbital inclination of approximately 63 deg for \psr\ from Tab.~\ref{table:timing_params} and leaving $\eta$ as a free parameter, we derive 
\begin{equation}
  |\dot\omega|_{\rm tidal,max} \approx 
  8.1\times 10^{-5}
  \left(\frac{m_3}{1\,\msun}\right)
  \left(\frac{R}{100\,\mathrm{AU}}\right)^{-3}
  \mathrm{deg\,yr}^{-1} \,.
\label{eq:maximum_omdot_tidal}
\end{equation}
Similarly, from equation (\ref{eq:Di3}), using $\Delta x = x\,\cot i \,\Delta i$, one finds
\begin{equation}
  |\dot{x}|_{\rm tidal,max} \approx 
  1.8\times 10^{-13}
  \left(\frac{m_3}{1\,\msun}\right)
  \left(\frac{R}{100\,\mathrm{AU}}\right)^{-3} \,.
\label{eq:maximum_xdot_tidal}
\end{equation}

On a final note, in the limit for small eccentricities, when neglecting terms $\mathcal{O}(e^2)$, Eqs.~(\ref{eq:De3}) and (\ref{eq:Di3}) agree with the corresponding results obtained by \cite{Rasio_1994} and used by \cite{Joshi_Rasio_1997}. This is not the case for Eq.~(\ref{eq:Dom3}). It appears that \cite{Rasio_1994} has overlooked the vertical (out of plane) tidal component (Eq.~4 in \citealt{Rasio_1994}) when calculating $\dot\omega$. Nevertheless, even for orbits with a low eccentricity, this contribution is relevant. It results (generally) in a tilting of the orbital plane, which leads to a change of the line of the nodes within the plane of the sky and, more importantly, within the orbital plane, which in turn contributes to $\dot\omega$, since $\omega$ is defined with respect to the ascending node. This effect gives rise to the `$-\cos i \, \Delta\Omega$' term in Eq.~(\ref{eq:Dom3}), which is absent in Eq.~(5) of \cite{Rasio_1994}. 

%%%%%%%%%%%%%%%%%%%%%%%%%%%%%%%%%%%%%%%%%%%%%%%%%%%%%%%%%%%%%%%%%%%%%%%%%%%%%%%%

\section{Modifications to orbital models}
\label{sec:DD_modification}

The large line-of-sight jerk for a pulsar in a GC is a measure of the dynamical influence from an external body, and this affects both the spin period and the orbital period of the pulsar. This is evident in the resulting measurement of the large second period derivative, and suggests considerable orbital variability in the system. In this scenario, the variations in the orbital period ($P_\mathrm{b}$) can be modelled with multiple orbital frequency derivatives, and the orbital frequency $f_\mathrm{b}$ is the rate of change of the the number of orbits. At any given time $t$, the number of orbits can be expressed in a Taylor expansion as 
\begin{equation}
    N_\mathrm{b} = \sum_{k=0}^{n}\frac{1}{(k+1)!} \, f_b^{(k)}(t-T_0)^{k+1} \,,
    \label{eq:Orbit_number}
\end{equation}
where n is the number of orbital frequency derivatives, and $f_\mathrm{b}^{(k)}$ is the $k^\mathrm{th}$ orbital frequency derivative at the reference epoch of periastron passage $T_0$. 

However, if we're confident that the ``intrinsic'' contributions to these higher spin and orbital frequency derivatives are negligible (for instance, when the expected contributions  to $\ddot{f}$ and $\dddot{f}$ from timing noise are much smaller than the observations),
the orbital frequency derivatives can be obtained from the higher order spin frequency derivatives according to eq.~\ref{eq:dfbn}. Replacing those in equation~(\ref{eq:Orbit_number}), and using the $P_\mathrm{b}$ and $\dot{P}_\mathrm{b}$ used in those models (PB and PBDOT), we obtain:
\begin{align}
& N_\mathrm{b} = \frac{1}{P_\mathrm{b}}
    \Bigg[ (t -T_0) -
    \frac{1}{2} \frac{\dot{P}_\mathrm{b}}{{P_\mathrm{b}}} (t-T_0)^{2} 
\nonumber\\ 
& \qquad\qquad +
    \frac{1}{6} \frac{\ddot{f}}{f}(t-T_0)^{3} +
    \frac{1}{24} \frac{\dddot{f}}{f}(t-T_0)^{4} + \dots
    \Bigg] \,, \label{1851A_Orbit}
\end{align}
where the reference epoch for the measurement of the $\dot{f}^{(b)}$ must be close to $T_0$. We have thus modified the  DD, DDK, DDH and DDGR orbital models, where the second part of the equation is used if a flag (JERK) is set to 1 and any higher-order spin frequency derivatives are being fit. These models then produce self-consistent values for $f^{(n)}$ (with $n > 1$), where the effect of the jerk is taken into account not only for the spin phase, but also for the orbital phase. If the JERK flag is set to 0, or not specified at all, then only the first part of the equation is used, as in the current implementation. 

\end{document}